\documentclass[manuscript]{aastex}

\newcommand{\coa}{$\rm ^{12}CO$}
\newcommand{\cob}{$\rm ^{13}CO$}
\newcommand{\coc}{$\rm C^{18}O$}

\newcommand{\ci}{[CI]}
\newcommand{\cii}{[CII]}
\newcommand{\hi}{HI}

\usepackage{etaremune}

\slugcomment{Accepted by The Astrophysical Journal, 13 August 2015}

\shorttitle{Extended Carbon Emission in G328}
\shortauthors{Burton et al.}

\begin{document}


\title{Extended Carbon Line Emission in the Galaxy:  \\ Searching for Dark Molecular Gas along the G328 Sightline}


\author{Michael G. Burton, Michael C.B. Ashley, Catherine Braiding, Matthew Freeman}
\affil{School of Physics, University of New South Wales, Sydney, NSW 2052, Australia}

\author{Craig Kulesa}
\affil{Steward Observatory, The University of Arizona, \\ 933 N. Cherry Ave., Tucson, AZ 85721, USA}

\author{Mark G. Wolfire}
\affil{Astronomy Department, University of Maryland, College Park, MD 20742, USA}

\author{David J. Hollenbach}
\affil{Carl Sagan Center, SETI Institute, \\ 189 Bernado Avenue, Mountain View, CA 94043--5203, USA}

\author{Gavin Rowell} \and  \author{James Lau}
\affil{School of Chemistry and Physics, University of Adelaide, Adelaide, SA 5005, Australia}




\begin{abstract}
We present spectral data cubes of the \ci\ 809\,GHz, \coa\ 115\,GHz, \cob\ 110\,GHz and HI 1.4\,GHz line emission from an $\sim 1$ square degree region along the $l = 328^{\circ}$ (\objectname{G328}) sightline in the Galactic Plane. 
Emission arises principally from gas in three spiral arm crossings along the sight line.  The distribution of the emission in the CO and \ci\ lines is found to be similar, with the \ci\ slightly more extended, and both are enveloped in extensive HI\@.  Spectral line ratios per voxel in the data cubes are found to be similar across the entire extent of the Galaxy. However, towards the edges of the molecular clouds the \ci/\cob\ and \coa/\cob\ line ratios rise by $\sim 50$\%, and the \ci/\hi\ ratio falls by $\sim 10$\%.  We attribute this to these sightlines passing predominantly through the surfaces of photodissociation regions (PDRs), where the carbon is found mainly as C or C$^+$, while the H$_2$ is mostly molecular, and the proportion of atomic gas also increases.   We undertake modelling of the PDR emission from low density molecular clouds excited by average interstellar radiation fields and cosmic-ray ionization to quantify this comparison, finding that depletion of sulfur and reduced PAH abundance is needed to match line fluxes and ratios.  Roughly one-third of the molecular gas along the sightline is found to be associated with this surface region, where the carbon is largely not to be found in CO\@.  
$\sim 10\%$ of the atomic hydrogen along the sightline is cold gas within PDRs.

\end{abstract}





\section{Introduction}
\label{sec:intro}
One of the basic activities of a spiral galaxy like our own Milky Way is the continual collection of diffuse and fragmented gas and dust clouds into giant clouds of molecules, which in turn produce stars \citep[e.g.][]{2014prpl.conf..125M}.  This takes place as part of a cycle of matter between the stars and the gas, driven by energy flows arising from radiation and mass loss from stars, and the dynamical motions of the gas in the gravitating interstellar medium.   Cloud formation occurs as the gas cools and the density rises, first as atomic clouds, then as molecular clouds.  Clouds coalesce, cooling is enhanced due to molecule formation as radiation is shielded from their interior, and star formation initiated as gravitational collapse is triggered inside.

Determination of the spatial structure and the kinematics of the gas as it transitions between the phases of the gas is needed to discern how clouds are formed.  While the hydrogen gas can be seen from the atomic phase through its principal 21\,cm emission line, its measurement does not distinguish between the warm and cold phases, only providing the total column density of the atomic gas.  In the molecular phase the bulk component -- hydrogen molecules -- cannot be seen at all, remaining unexcited in their ground state at the typical $10-20$\,K temperatures found in molecular clouds.

Trace species in the gas are thus needed with emission lines that are sensitive to the excitation conditions in the gas, to follow the thermal and chemical transitions that occur.  After helium (which is inert) and oxygen (whose key lines are blocked by the Earth's atmosphere), carbon is the next most abundant element in the Universe.  Carbon can be found in ionized (C$^+$), atomic (C) and molecular (CO -- carbon monoxide) forms in the bulk of the dense interstellar medium.  All of these species are readily excited in the prevailing conditions, with emission lines produced in the terahertz portion of the spectrum for C and C$^+$ ($0.5 - 1.9$\,THz), and the millimeter portion for CO (3\,mm $\equiv 0.1$~THz).

CO is readily measured from good observatory sites.  C and C$^+$ are, however, virtually unobservable from all but the driest sites on the surface of our planet.  We have established a new observatory at Ridge A, near the summit of the Antarctic plateau, to open up the terahertz spectrum for observation in order to obtain the wide-field, high resolution images of the carbon lines needed to pursue this science \citep{2012SPIE.8444E..1RA, 2013IAUS..288..256K}.  The measurement of the key diagnostic lines from all these species in the atomic and molecular phases of interstellar medium we call ``following the galactic carbon trail''.  

In this paper we present wide-field data from the Mopra telescope in Australia, obtained as part of the Southern Galactic Plane CO survey \citep{2013PASA...30...44B, 2015PASA...32...20B} and the HEAT (High Elevation Antarctic Terahertz) telescope at Ridge A in Antarctica of atomic carbon, aimed at pursuing this objective. The data covers roughly 1 square degree along the G328 sightline through the Galactic plane (i.e.\ $l = 328^{\circ}$).  In particular, we have examined the data set to determine whether there is evidence for dark molecular gas present.  This is defined here as regions exhibiting \ci\ emission but without corresponding CO line emission.  Practically, since there are no clear regions where \ci\ is present but CO is not detected in the moderate beam sizes used (2 arcmin for the \ci), we search for regions of emission where the [C/CO]  abundance may be enhanced.

Such regions of gas may be expected in the surfaces of molecular clouds.  These are photodissociation regions \citep[PDRs; e.g.][]{1985ApJ...291..722T, 1990ApJ...365..620B, 1992ApJ...399..563B, 1993ApJ...402..195W}, and within extinctions of $A_V \lesssim 1$\,mags.\ from the atomic surface, self-shielding may allow significant columns of H$_2$ to exist.  However the CO abundance will be greatly reduced from the cloud interior, being photodissociated by the far-UV radiation that heats and drives the chemistry inside the PDR\@.  Since CO is the normal tracer used to indicate the presence of molecular gas, such gas is ``dark'' to standard survey techniques.  In these molecular cloud envelopes the carbon will instead be found as either C or C$^+$, and so be amenable to detection through THz frequency observations.  

\citet{2010ApJ...716.1191W} model the fraction of such a dark component that exists in giant molecular clouds, suggesting that it comprises about one-third of the molecular gas, in good agreement with the estimates of the dark gas fraction from gamma-ray observations \citep{2005Sci...307.1292G}.  In these and previous cloud models  \citep[e.g.][]{1988ApJ...334..771V} the \ci\ emission arises mostly from gas where the hydrogen exists as H$_2$ but without significant CO present. Thus \ci\ line emission can be used to trace the dark molecular gas.  

Furthermore, small molecular clouds, here defined as those with total column densities such that $A_V \lesssim 1\, {\rm mag.} \equiv {\rm N_{H_2}} = 10^{21} {\rm cm}^{-2}$, would not be found at all in CO surveys. A population of such clouds could remain unseen without a corresponding carbon-line survey to detect their presence.  

There is also the possibility of dark atomic gas, where 21\,cm HI is optically thick and so its line intensity under-estimates the gas column.  Analysis of the Planck satellite data \citep{2011A&A...536A..19P}, comparing the optical depth of the  dust emission at 353\,GHz to the column of the HI derived from its 21\,cm line emission,  suggests that such clouds may be widespread across the Galaxy.  This can occur when the atomic gas is cold ($T_S < 80$\,K), through absorption of background continuum.  \citet{2015ApJ...798....6F} examine this further for atomic clouds within a few hundred parsecs of the Sun, but out of the Galactic plane.  They suggest that, if the column density derived from the dust is indeed all atomic,  then there are  extensive regions where $T_S$ is below 35\,K, with optical depths $\tau_{\rm HI}$ as high as 3.0; i.e.\ dark atomic gas\footnote{We note that \citet{Lee2015} argue that these results are incorrect and that the HI is not optically thick in the Perseus region that they investigated, though the two studies do probe the ISM on very different spatial scales.}.   Carbon would exist as either C or C$^+$ in such atomic clouds (but not as CO), and so emission from the THz frequency lines of these species might be used as a tracer of such gas.  This is especially so in the Galactic plane where there is also extensive molecular gas, making it difficult to separate out the dust associated with the atomic component.

By mapping the distribution of carbon along the Galactic plane we can therefore estimate where the dark gas may be distributed and how much of it exists within the interstellar medium.  In this paper we concentrate on an analysis of the carbon emission associated with the molecular gas on the Galactic sightline that passes through the G328 sector, also comparing it the \coa, \cob, \coc\ and HI line emission.  We describe the observations in \S\ref{sec:obs} and present the data in \S\ref{sec:results}.  Examining the variation of [C/CO] at the edges of emission features, where the fluxes are smallest, presents several challenges so we undertake the analysis through a variety of techniques: flux images (\S\ref{sec:moment}), histograms of line ratio distributions (\S\ref{sec:histograms}), averaged line profiles (\S\ref{sec:profiles}) and a relatively new technique: saturation-hue plots (\S\ref{sec:evans}).  We undertake PDR modelling for the molecular environment appropriate for the bulk of the data set in \S\ref{sec:pdr}, presenting several Figures showing line fluxes, optical depths and line ratios as a function of depth into the PDR in \S\ref{sec:pdrmodelplots}.  We then compare these to the data in \S\ref{sec:discuss} and summarize the results in \S\ref{sec:summary}.  This is Paper II.  Paper I \citep{2014ApJ...782...72B} analyzed this same data set, but just for a narrow velocity range associated with a single molecular filament.

\section{Observations and Data Reduction}
\label{sec:obs}
The data analyzed here comes from three separate spectral line surveys conducted along the southern Galactic plane, of THz-band \ci, millimetre-band CO and centimeter-band HI line emission.  The 62\,cm HEAT (High Elevation Antarctic Terahertz) telescope has imaged the \ci\ 809.3\,GHz line ($\rm ^3P_2 -  ^3P_1$) from Ridge A, Antarctica, yielding data cubes with $2'$ and 0.7\,km/s resolution \citep{2013IAUS..288..256K}.  \coa, \cob\ and \coc\  J=1--0 emission lines (115.3, 110.2 \& 109.8\,GHz, respectively) were imaged as part of the 22\,m Mopra telescope Southern Galactic Plane CO Survey \citep{2013PASA...30...44B, 2015PASA...32...20B}, with $0.6'$ and 0.1\,km/s resolution.  The 1.420\,GHz HI line comes from archival data in the Parkes--ATCA Southern Galactic Plane Survey \citep[the SGPS;][]{2005ApJS..158..178M} and provides $2'$ and 3\,km/s resolution. We will generally refer to these lines as simply \ci, \coa, \cob, \coc\ \& HI, respectively,  in this paper, though sometimes will refer to the \ci\ line as \ci\ 2--1 to distinguish it from the [CI] $\rm ^3P_1 -  ^3P_0$ (1--0) line at 492\,GHz.  We will also include discussion of the \cii\ $\rm ^3P_{3/2} -  ^3P_{1/2}$ 1.90\,THz line, referred to here simply as \cii\ (and also commonly known as the 158$\mu$m line).

The contiguous data set obtained for the C, CO and H lines shown here covers $l=327.8$--$328.7^{\circ}, b=-0.4$ to $+0.4^{\circ}$.  This data set was first presented in \citet[][hereafter Paper I]{2014ApJ...782...72B}, where an analysis of the emission from a filamentary feature, approximately 75\,pc long and 5\,pc wide, and with only a narrow 2~km/s FWHM in all these spectral lines, was presented, hypothesized to represent a giant molecular cloud in the process of formation.  In this paper we now examine the data from the complete range of velocities covered by these observations with the HEAT telescope, taking in all of the emission from across the Galaxy along this sight line.

The Mopra CO and SGPS HI cubes were re-binned to the same voxel resolution as the HEAT CI cubes (both in position and velocity; i.e.\ $2'$ and 0.7\,km/s).  All cubes were then smoothed with a $2'$ FWHM Gaussian.   The analysis presented here concentrates on examining variations seen in the \ci/\cob\ line ratio in the data cubes.  The \cob\ line is used in preference to the brighter \coa\ line in order to minimize complications in interpretation caused by optical depth in the \coa\ line, as the \cob\ line is generally optically thin in the Mopra CO survey \citep[see][] {2013PASA...30...44B}.

The full velocity range of the [CI] emission extends from $-120$ to 0 km/s. Three principal features are seen in its averaged spectral profile (see Fig.~\ref{fig:profiles}, where the \ci\ line is shown together with the CO and HI profiles). Fig.~\ref{fig:galaxymodel} shows a schematic model of the Galaxy \citep[adapted from][]{2014AJ....148....5V} to aid in the interpretation of these profiles.  The three features correspond to spiral arm crossings at $-100$ to $-85$~km/s, $-80$ to $-65$~km/s and $-55$ to $-40$~km/s, respectively.  In turn, they equate predominantly to emission from the Norma spiral arm near- and far- crossings, and the Scutum-Crux near-crossing along the G328 sight line, at distances of $\sim$6, 10 \& 3.5\,kpc from Sun.   In addition, the velocity range from $-80$ to $-76$~km/s  corresponds to the quiescent filament analyzed in Paper I, at $\sim 5$\,kpc distance.  The data set is split into these four velocity ranges for the analysis described below. There is also a weaker feature at $-20$~km/s arising from the Scutum-Crux far-arm crossing, $\sim 13$\,kpc distant. 

A visual rendering of the data cubes for G328 in the four principal lines studied here (HI, \ci, \cob, \coa) is also shown in Fig.~\ref{fig:3dview}.  Here the velocity dimension is extended along the long axis, and provides a proxy for distance from the Sun (with the proviso of the near-far ambiguity).  The three features corresponding to the spiral arm crossings are readily apparent (Norma near-, Norma far- and Scutum-Crux near-, going from left to right).  The greater extent of the atomic hydrogen is clear, enveloping the carbon and carbon monoxide emitting gas.  The carbon also gives the impression of being more extended than the carbon monoxide,  in-between the distributions for the atomic and the fully molecular gas.  However care needs to be given towards making such an interpretation, as the relative optical depths of the emitting species, as well as the display scales chosen, can result in mis-leading visual appearances.  In the rest of this Paper we concentrate on a quantitative analysis of the relative distributions of the hydrogen, carbon and carbon monoxide emission to ascertain whether the impression given here is correct, and to explain why and when such a separation can occur.

\section{Results}
\label{sec:results}
In this section we first consider the integrated flux images for the CO and \ci\ lines (i.e.\ zeroth moment images) in order to compare their morphologies in each of the four velocity ranges.  While these show suggestions of a more extended distribution for the \ci\ emission than the CO this is not conclusive, so we then consider other measures that can probe their relative distributions.  This includes histograms of the distribution of lines ratios and filtering the data to examine averaged profiles from voxels with different values of the \ci/CO line ratio.  We also present a new method of presentation (Evans plots) which applies a 2D color table (hue and saturation) in order to visualize two variables of interest in such a way that pixels with low signal to noise do not dominate a display of the ratio of two intensities.

\subsection{Moment Images of Line Flux Distributions}
\label{sec:moment}
Integrated flux (Moment 0) images over the four velocity ranges are shown in Fig.~\ref{fig:momentimages120} for both the \cob\ and \ci\  lines, accompanied by an image of the \cob\ overlaid with \ci\ contours.  These images trace the locations of the molecular clouds within three spiral arm crossings along the G328 sight line (Norma far-, Norma near- and Scutum-Crux near-), and so are all quite different.   The head of the filamentary structure studied in Paper I is to the left of centre of the lowest panel in this Figure ($\sim 328.4^{\circ}, -0.1^{\circ}$)\footnote{The full filament can be seen in CO and HI in Fig.~1 of Paper I.}.

The \cob\ and \ci\ images clearly look similar in each of the velocity ranges.  However their overlay in the third column suggests that the \ci\ is, in general, smoother and more extended than in the \cob\ images.  To check that this conclusion is not an artifact of the different original resolutions of the \cob\ and \ci\ data sets the analysis was also repeated with the cubes smoothed by a $200''$ Gaussian rather than $120''$.  The same result still holds (though is not shown here), of the \ci\ being smoother and more extended than the \cob, though it is not especially striking when seen at this lower angular resolution.

This is as expected if the carbon is wrapped around the molecular gas, as well as being contained within it.  Such behaviour would occur if the outer portion of each molecular cloud was lacking in CO\@.  However, while suggestive this result cannot be regarded as definitive from these images, so we now examine other metrics to probe the relative distributions of the \ci\ and CO emitting gas.

\subsection{Histograms of Line Ratio Distributions}
\label{sec:histograms}
We show in Fig.~\ref{fig:histograms} histograms of selected ratios from the \ci, \coa, \cob, \coc\ and HI lines in the data.  These distributions are determined per voxel (i.e.\ per $2'$ spatial pixel,  0.7 km/s velocity channel).  To be included here the data for a \ci\ voxel needed to be $> 3\sigma$, whereas for the three CO lines we show histograms for thresholds of  $1\sigma$ and $3\sigma$.   These $\sigma$ values are determined from the standard deviation of voxels in the continuum portion of the data cubes, $-120$ to $-100$\,km/s.  For the HI data, for which $\sigma$ is not well defined since there are no clear regions of line-free emission in the spectra, but whose signal-to-noise (S/N) is much higher than that of the \ci\ and CO, the threshold was set to an arbitrary value.  The number of voxels passing these threshold criteria is listed in Table~\ref{tab:voxels}, and some statistics on the resulting histograms in Fig.~\ref{fig:histograms} are listed in Table~\ref{tab:voxelstats}.  These are given for both the $1 \sigma$ and $3 \sigma$ thresholds applied to the CO fluxes.  The mean values and standard deviations for the ratios \coa/\ci, \cob/\ci\ and \coa/\cob\ are $6.8 \pm 1.6, 1.4 \pm 0.5 \, {\rm and} \, 5.5 \pm 2.5$, respectively, when the $1 \sigma$ threshold for CO is applied.

The intent here is to search for regions of strong \ci\ and weak (or absent) CO emission,  hence the $1\sigma$ threshold limit applied to the CO data.  There is limited dynamic range in the data set, so repeating this analysis with a $3\sigma$ cut-off for CO yields few voxels where the \ci/\cob\ ratio is ``high'' (as defined below).  We do, however, list metrics in Table~\ref{tab:voxelstats} for the corresponding distributions when a $3 \sigma$ threshold for the CO lines is applied.  The analysis discussed below yields consistent results with the $1 \sigma$ threshold when this is done, though from far fewer data points.

For \coa/\ci\ the distribution is roughly Gaussian, however for all the other line ratios shown in Fig.~\ref{fig:histograms} this is not the case.  A significant non-Gaussian tail exists for larger values of the other line ratios.  This highlights excess voxels where the line ratio is typically about twice the mean value.  The tail is especially evident for the \ci/\cob\ ratio, which has a mean value of $\sim 0.8$ and ``high'' values extending to $\sim 2$.  That the tail is also sensitive to the value of the threshold value used to select the CO lines is also apparent from the (dotted line) overlays in Fig.~\ref{fig:histograms}, where the corresponding distribution for \ci/\cob\ for the $3 \sigma$ threshold is almost Gaussian in form.  This, of course, means that care is needed in interpreting the data in the tail, as noise fluctuations from voxels with weak fluxes will cause some \cob\ values to be ``low'', and hence the corresponding \ci/\cob\ values to be ``high''.  Nevertheless, by averaging over voxels where the line ratios are high the S/N can be improved, allowing us to examine where such voxels tend to be distributed.  We undertake such an analysis in the following section (\S\ref{sec:profiles}).

Fig.~\ref{fig:ratiovsflux} presents a series of scatter plots showing the relationship between the line ratios (\ci/\cob, \ci/\coa\ and \coa/\cob)  and the fluxes of the CO and \ci\ lines, for every voxel in the datacube, where the $3 \sigma$ threshold is applied for the \ci\ flux, and $1 \sigma$ for \coa\ and \cob.  The top panels show the line ratios as a function of the \coa\ and \cob\ fluxes, the bottom line as a function of \ci\ flux.  As these fluxes increase, the line ratios tend towards roughly constant values, given by \ci/\cob$\sim 0.5$, \ci/\coa$\sim 0.15$ and \coa/\cob$\sim 3$. In contrast, as the fluxes decrease, the typical values of these line ratios tend to increase.  Of course, care needs to be given to such an interpretation, for noise fluctuations will result in higher derived line ratios for the lowest fluxes when the flux threshold applied is also the denominator in the ratio (as it is for the plots in the top line).\footnote{Note also that the curvature seen here as the lower boundary in the first two plots in Fig.~\ref{fig:ratiovsflux} represents the ratio when the threshold value is applied to the numerator; i.e.\ the \ci\ line.}  Nevertheless, the median values of the \ci/CO lines ratios  are clearly rising as the CO line fluxes decrease.  The overplotted lines in Fig.~\ref{fig:ratiovsflux}  show the weighted minimum least squares linear fits to the data\footnote{The weighted fit parameters are listed in the caption to Fig.~\ref{fig:ratiovsflux}.  These linear fits are strongly weighted towards lower ratios for the lower fluxes as the S/N is highest for such data points.  Hence the fit lines fall below the bulk of the data points.}.  The weak, but clearly negative slopes, to the fits plotted against the CO flux, as well as to \ci/\cob\ vs.\ \ci, quantify this relationship; e.g.\  when the \cob\ flux is 1\,K in a voxel, the \ci/\cob\ ratio is $\sim 50\%$ higher than when it is 3\,K, and for a \ci\ flux of 0.5\,K in a voxel, the \ci/\cob\ ratio is $\sim 40\%$ higher than when it is 1.5\,K\@.

The \cob\ line flux, being generally optically thin, provides a measure of the product $N_{\rm CO} f_{\rm CO}$, the column density of the CO molecules times their filling factor in the beam, whereas for the \ci\ flux it is $N_{\rm C} f_{\rm C}$\footnote{Note that we are implicitly assuming that the \ci\ emission is also optically thin.  This is consistent with the results of the PDR models we present later, and was also the case for the analysis given in Paper I.}. While for any given voxel these factors cannot be separated, statistically, as the flux decreases there will be a greater proportion of lower column density regions associated with the corresponding voxels.  We deduce that for these voxels the ratio of the columns of gas in C to CO tends to rise, although the total column in both species is lower.  

\subsection{Averaged Line Profiles}
\label{sec:profiles}
We pick a dividing ratio between ``high'' and ``normal'' of 1.0 for \ci/\cob\ from inspection of the histogram in Fig.~\ref{fig:histograms}, and then split the data set into two cubes for those voxels that are ``high'' and those that are ``normal''.  Visual inspection of the resulting cubes suggests that ``high'' ratio values are generally found around the edges of CO emission regions.  However, care is needed over any interpretation of high \ci/\cob\ existing at the peripheries of molecular clouds because these are also the regions where the S/N is lowest since the flux is least. In particular, the $1\sigma$ threshold limit applied to the selection of CO voxels exacerbates the selection of such voxels at the edges of emitting clouds. 

\subsubsection{Splitting the Data based on the \ci/\cob\ Line Ratio}
\label{sec:split}
We thus apply a mask to separate the voxels with ``high'' and ``normal'' ratios, and average the resulting cubes to improve the resulting S/N (data from $\sim19,000$ and $\sim55,000$ voxels are averaged, respectively).  The results are displayed in Fig.~\ref{fig:avprofiles}.  Here are shown the average profiles for \cob\ (in red) and \ci\ (green) in the data cube for all voxels that have ``high'' and ``normal'' \ci/\cob\ ratios (i.e.\ $>$ and $< 1.0$, respectively, as well as passing the flux thresholds discussed in \S\ref{sec:histograms}).  Profiles for voxels with ``normal'' ratios are drawn with the solid lines, while those for ``high'' \ci/\cob\ ratios are shown with the dotted lines.  It is clear  that ``normal'' ratios generally correspond to brighter fluxes for both \ci\ and \cob, as well as having the \cob\ line flux about 50\% stronger than the \ci\ flux (i.e.\ red $>$ green).  Profiles for ``high'' \ci/\cob\ voxels, on the other hand, show weaker fluxes on average, as well as (by design) \ci\ fluxes that are (slightly) higher than \cob.  This result applies across all velocities, and in particular in each of the three principal spiral arm crossings along the G328 sightline.  Since the resulting profiles have been averaged across hundreds of voxels meeting the selection criteria their S/N is good.   We can conclude that \ci/\cob\ is higher in pixels where the \ci\ and \cob\ line fluxes are lowest; i.e.\ as would occur preferentially on the edges of molecular clouds.

There is also the possibility that high ratios occur in voxels which have low optical depth (or column density), for, as we discuss in \S\ref{sec:pdr}, clouds which only extend for $A_v \sim 1-2$\,mags.\ have a large proportion of their gas in the surface layer (i.e.\ $A_v < 1$). Here the carbon will predominantly be found as C--atoms (both singly-ionised and neutral).  However, as the \coa/\cob\ line ratios shown in Fig.~\ref{fig:ratiovsflux} illustrate, their typical flux ratio is $\sim 5$ and virtually all voxels have [\coa/\cob] $< 10$; i.e.\ the \coa\ remains strongly optically thick.\footnote{$\tau \sim 5$ when [\coa/\cob] = 10, for an isotope ratio [$^{12}$C/$^{13}$C] of 50.} Thus, this possibility (low optical depth) could only apply to relatively few voxels in the data set, at most.

\subsubsection{Line Profiles and Ratios as a Function of Velocity}
We now examine the CO, \ci\ and \hi\  lines profiles and ratios as a function of velocity, comparing the two groups (``normal'' and ``high'' [CI]/$^{13}$CO ratios) with each other.

Fig.~\ref{fig:lineprofiles} shows the mean profiles for the lines, and Fig.~\ref{fig:lineratios}  for their line ratios, when the data has been selected so that the \ci/\cob\  ratio is ``high'' (i.e.\ $> 1.0$) or ``normal'' (i.e.\ $< 1.0$) (as well as thresholded, as before\footnote{The profiles have been constructed for each voxel meeting the criteria, so the normalization is different for each velocity channel; hence the different appearance from Fig.~\ref{fig:avprofiles}.}). The behaviour is similar for the three velocity ranges (i.e.\ spiral arm crossings).  The \coa\ and \ci\  fluxes are slightly brighter, on average, when the \ci/\cob\ ratio is ``normal'',  whereas the \cob\ fluxes are significantly stronger for ``normal'' ratios.  The \hi\ flux is, however, is slightly smaller for ``normal'' \ci/\cob\  ratios. 

For the plots showing line ratios (Fig.~\ref{fig:lineratios}),  the \ci/\cob\ ratio is significantly larger for the ``high''  voxels (as it should be, by design).  The \coa/\cob\ ratio is also similarly higher. There is, however, little difference in the value of the \ci/\coa\  ratio between the ``high'' and the ``normal'' voxels. None of these line ratios depend on the velocity, either.  They are constant, only dependent  on the relative value of \cob\ line flux; i.e.\ higher \ci/\cob\ and \coa/\cob\ ratios for voxels with lower \cob\ fluxes, and vice-versa.  

The \ci/\hi\ ratio, however, {\it decreases} going from ``normal'' to ``high'' \ci/\cob\ values, by about $\sim 10-20\%$.  This indicates that there is slightly more atomic gas associated with the ``high'' \ci/\cob\ ratio gas.  Also noticeable is the velocity range for the narrow, quiescent filament ($-80$ to $-76$\,km/s) discussed in Paper I, where the \ci/\hi\ ratio is about 40\% higher where \ci/\cob\ is ``normal'', indicating substantively more atomic gas associated with this filament than for other molecular gas.

The conclusion is that high \ci/\cob\ and \coa/\cob\ ratios occur when the \cob\ flux is weak -- i.e.\ on the edges of the various molecular emission features.  At these same locations the \coa\ emission is not, however, significantly weaker, on average.  This can be attributed to optical depth; the \coa\ line remains optically thick, even when relatively less of the species is present, whereas the same sightline samples all of the \cob\ in the gas, if it passes through the edges of clouds.  Since the \cob\  flux is lower here then, as all the CO is being sampled, its total column must also be lower.  In these same sightlines we find that the \ci\ flux is only slightly lower, however, and hence so is the column of carbon.  Thus, elevated carbon to carbon monoxide column density ratios are measured towards the edges of all emission features, i.e.\ at the edges of molecular clouds.  

Furthermore, the lower \ci/\hi\ ratios in this same gas indicates that this material is associated with relatively more atomic gas at molecular cloud edges. This is as expected for sightlines passing preferentially through surface layers ($A_V < 1$) of PDRs, where CO should be photodissociated into C and C$^+$ and a significant amount of H$_2$ is photodissociated into H\@.  

Thus, the relative column of C to CO along these sightlines is increased in comparison to sightlines passing through the interiors of molecular clouds. This is also the observational signature we have defined for detecting ``dark gas''.  This amounts to a statistical detection of regions where the carbon (C) is enhanced in abundance compared to the carbon monoxide (CO).  These must preferentially contain the surface regions of molecular clouds since they pass through their edges.  Finally, for the quiescent filament, which we hypothesized to be undergoing the process of molecular cloud formation in Paper I, the increased \hi\ abundance associated with the cloud edges is consistent with the assertion that it is condensing out of the atomic substrate.

\subsection{Hue -- Saturation Plots to Examine Variations in Line Ratios}
\label{sec:evans}
We present here another technique to examine the significance of the increased \ci/\cob\ ratios at the edges of the molecular clouds -- hue--saturation plots.

Hue / Saturation plots\footnote{These are sometimes known as Evans plots; see the description given at http://www.ncl.ucar.edu/Applications/evans.shtml.} provide a way of visualizing two variables of interest, one of which provides some measure of ``importance'', and the other its value.  This is done by assigning the hue of a color to the value  and the saturation (or intensity)  of the color to the ``importance''.   Hue is cyclic, representing the angle around a color wheel running through the spectrum from red to purple (and back to red again); i.e.\ with range from $0^{\circ}$ up to $360^{\circ}$.  Low importance causes the color to fade into grey regardless of what the value actually is.  Such plots are useful for examining the ratio of two variables when the overall S/N is limited.  It avoids the eye being drawn to seemingly high line ratios in regions of low S/N data.

We have constructed such Evans plots to examine the variation of the line ratios by using hue (i.e.\ color) to denote the value of the line ratio, and saturation (i.e.\ intensity) to denote its S/N.  Hue is ranged from 0 to $300^{\circ}$ in these plots to avoid wrapping, i.e.\ from red for the lowest ratio displayed to purple for the highest.  Saturation is ranged from 0 to 1, with 1 normalized to the value of the highest S/N.  Hue-Saturation-Lightness values (the latter, which represents the perceived luminance of the system, is always set to a fixed value of 0.5 here) are then converted to red-green-blue for display (using IDL library routines which transform between the HSL and RGB color systems).

Two Evans plots are shown in Figs.~\ref{fig:evans_v8076} -- \ref{fig:evans_v100} for the velocity ranges of $\rm (-80,-76)$ and  $(-100,-85)$~km/s; i.e. the filament and the Norma near- spiral arm crossing (the other two spiral arm velocity ranges show similar characteristics and are included in the Appendix; see  Figs.~\ref{fig:evans_v80} -- \ref{fig:evans_v55}).   For each velocity range three plots are displayed, \ci/\cob, \ci/\coa\ and \coa/\cob, so the behaviour of all three lines can be compared.   The Evans plots are on the left and their corresponding 2D color tables on the right.

Examination of these images (and in comparison to the flux images of Fig.~\ref{fig:momentimages120}) shows that:

\begin{enumerate}
\renewcommand{\theenumi}{\roman{enumi}}
\item For \ci/\cob,  the highest S/N regions tend to be surrounded by higher-ratio, but lower S/N regions.

\item For \ci/\coa, on the other hand, there are regions of differing line ratio across the fields, but with no particular tendency for high ratio regions to surround low ratio regions, or vice-versa. There is also no tendency for lower S/N regions to have different ratios than higher S/N regions.  

\item For \coa/\cob, the line ratios are generally lower within the emission regions where there is more CO, and higher at their edges.

\end{enumerate}

Taking these results in reverse order, we can interpret them as follows:

\begin{etaremune}
\renewcommand{\theenumi}{\roman{enumi}}
\item Since \coa/\cob\  variations measure optical depth changes, we deduce that the optical depth is highest (and so the line ratio is lowest) along the sight lines passing through cores of clouds.  It is lowest (with the highest ratios) for sight lines that pass through the edges of clouds.

\item \ci/\coa\  ratio variations trace changes in the relative \ci\ flux between molecular clouds, since \coa\ is, in general, optically thick and so relatively constant.

\item Higher line ratios for \ci/\cob\ are seen for sight lines that do not pass through the cores of GMCs, but rather tend to pass through their peripheries.  Since this ratio is sensitive to all the C and CO gas along each sight line, as both lines are largely optically thin, then along them the C abundance must be elevated compared to that of the CO\@.  This is as expected for the surface layers of molecular clouds, where for the first optical depth CO will generally be photodissociated, and so the carbon can only be seen as \ci\ (and \cii), rather than as CO\@.  This is likely mostly dark molecular gas, with $\rm H_2$ existing in the surface layer, and the carbon largely in atomic form rather than molecular.  

In the PDR models discussed in the following section the \ci\ peaks in gas which is molecular rather than atomic.  It may, however, also represent atomic gas, in particular if there is \cii\ present.  As determined in Paper I, the column density of \hi\ is typically $\rm \sim 10^{21}\,cm^{-2}$ in each voxel.  Future measurements, at the same resolution, of the 1.9 THz \cii\ line, which will co-exist with much of the \hi, would allow the amount of C$^+$ and C associated with the atomic gas to be quantified.
\end{etaremune}

There are also specific regions which stand out in these plots.   For instance, at $(328.65,-0.25)$ in Fig.~\ref{fig:evans_v8076} both \ci/\cob\ and \coa/\cob\ are measured to be about twice as high as other regions within the filament, while \ci/\coa\ is unchanged.  This can be ascribed to a correspondingly lower optical depth in the CO\@.

\section{Photo Dissociation Region (PDR) Models}
\label{sec:pdr}

We have examined a range of cloud properties using a modified form of the \cite{2010ApJ...716.1191W} photodissociation region code to calculate column densities and line emission to compare with the data.  The model consists of a slab of gas of total optical depth $A_{V,{\rm tot}}$ illuminated by the interstellar radiation field on two sides. The gas temperature and the abundances of atomic/molecular species are calculated as a function of optical depth, $A_V$, under the assumptions of thermal balance, chemical equilibrium and constant thermal pressure. For details on the chemistry and thermal processes we refer to \citet{1985ApJ...291..722T}, \citet{2006ApJ...644..283K}, \cite{2010ApJ...716.1191W} and \cite{2012ApJ...754..105H}.

We have made a number of modifications to the code used in \cite{2014ApJ...782...72B}. First we have explicitly included ${\rm ^{13}C}$ chemistry and line transfer in order to model the \cob\ observations. A number of researchers have presented their results on ${\rm ^{13}C}$ chemistry in PDRs including \cite{2009A&A...503..323V}, \cite{2013A&A...550A..56R} and \cite{2014MNRAS.445.4055S}. Here we use the fit by \cite{2007A&A...476..291L} to the temperature dependence of the  important fractionation reaction ${\rm ^{13}C^+} + {\rm ^{12}CO}\rightarrow {\rm ^{12}C^+} + {\rm ^{13}CO} +34.8 {\rm K}$, and use a ${\rm ^{12}C/^{13}C}$ ratio of 50 as appropriate for the inner Galaxy \citep{2005ApJ...634.1126M}. Second, we found that a deeper penetration of the external radiation field, compared to our previous models, is required to accurately fit the \ci/\cob\ line ratios. Similar results have been reported by \cite{2015MNRAS.448.1607G} and naturally arise in a turbulent gas in which lines of sight pass through low column density material. We approximate this process in our models by changing the angle dependence of the incident radiation field from isotropic to normally incident. For a gas layer at optical depth $A_V$, the resulting FUV field strength is $e^{1.8 A_V}$ times larger for the normally incident case compared to the isotropic case \citep[see discussion in][]{2010ApJ...716.1191W}.  We used the shielding functions for \coa\ and \cob\ given by \citet{2009A&A...503..323V} which accounts for self-shielding, shielding of \coa\ and \cob\ by H$_2$, and shielding of \cob\ by \coa.
 
Additional changes involve abundances of metals and Polycyclic Aromatic Hydrocarbons (PAHs). \cite{1985ApJ...291..722T} noted that an important  production reaction for C was charge exchange with ${\rm S}$ (${\rm C^+ + S \rightarrow C + S^+}$). In addition, it has since been realized that ion recombination on PAHs can be important in increasing the abundance of neutrals (and also decreasing the abundance of free electrons). We find that a decrease in both the sulfur abundance and PAH abundance is required to match the \cob\ to \ci\ line ratio. Our previous models used a sulfur abundance of ${\rm S/H} = 2.8\times 10^{-5}$ based on the analysis of \cite{1996ARA&A..34..279S}. However, a comprehensive study by \cite{2009ApJ...700.1299J} finds a sulfur abundance in depleted lines of sight in the solar neighborhood to be ${\rm S/H} = 3.5\times 10^{-6}$. In a recent study of the \ci\ line emission towards the Taurus molecular Cloud, \cite{2014ApJ...795...26O}  find a sulfur depletion factor of at least 50 is required to match the observations. In light of the more direct observational estimate of the S abundance and the likely presence of PAHs, we limit the sulfur depletion to the \cite{2009ApJ...700.1299J} value. Since we are modelling clouds in the inner Galaxy we multiply by a factor of two all gas phase metal abundances, dust abundances and PAH abundances to account for the increased metallicity there. This results in a gas phase sulfur abundance of $\rm S/H = 7\times10^{-6}$ (see Table~\ref{tab:pdrmodel}). 
 
Even using the depleted sulfur abundance we find that we still need to reduce the Galactic PAH abundance by a factor of two. Reducing the PAH abundance decreases the rate of $\rm S^+$ recombination on PAHs and thus decreases the S abundance in the $\rm C^+/C$ zone. This further reduces the neutral carbon abundance. \citet{2012ApJ...754..105H} used $\rm PAH/H = 2\times10^{-7}$ for a local abundance. Accounting for higher metallicities in the inner Galaxy would result in $\rm PAH/H = 4\times10^{-7}$, however we reduce this by a factor of two to arrive at a model abundance of $\rm PAH/H = 2\times10^{-7}$. There is still considerable uncertainty in both PAH rates and PAH abundances and a factor of two decrease is not unreasonable.

Our ``standard'' model parameters are listed in Table~\ref{tab:pdrmodel}, with some predicted lines fluxes and column densities then listed in Table~\ref{tab:pdrmodeloutput}.  For this standard model we adopt a radiation field of $G_0 = 5.1$  (with half of this incident on each side; $G_0$ is the interstellar field strength in units of Habing fields with $G_0 = 1.7$ representing an average field strength in the ISM; \citet{1968BAN....19..421H, 1978ApJS...36..595D}).   This field is roughly consistent with the interstellar field expected at a Galactocentric radius of 5\,kpc \citep{2003ApJ...587..278W}.  We also use a constant thermal pressure of $P_{\rm th}/k=2\times 10^4$ K ${\rm cm^{-3}}$.  The typical density (when $T_{\rm gas} \sim 30$\,K) is then $n_{\rm H_2} \sim 600$\,cm$^{-3}$. Models are computed as a function of total optical extinction through the cloud from $A_{V,{\rm tot}}=0.5$~mags.\ for `thin' clouds, up to $A_{V,{\rm tot}}=6$~mags.,  where $A_{V,{\rm tot}}=[N({\rm H~I})+2N({\rm H_2})]/2.0\times 10^{21}$ ${\rm cm^{-2}}$. We demonstrate in \S\ref{sec:pdrmodelplots} the effects of varying our standard  parameters.

We have adopted a primary cosmic ray ionization rate of  $\rm \zeta _{crp}=2\times 10^{-16}\,s^{-1}$, consistent with rates for low column density diffuse clouds determined from observations of H$_3^+$ \citep{2012ApJ...745...91I}, OH$^+$ and H$_3$O$^+$ \citep{2012ApJ...754..105H, 2015ApJ...800...40I}.    Furthermore, our modelling of the MeV--GeV cosmic-ray fluxes \citep[based on][]{2012A&A...541A.126S} from the known supernova remnants (G329.7+00.4, G327.4+01.0, G327.2-00.1, G327.1-01.1, G328.4+00.2 and G327.4+00.4) and Fermi-LAT GeV gamma-ray sources (3FGL J1554.4-5315c and 3FGL J1552.8-5330) in the region suggests no enhancement of the ionization rate beyond the galactic average (i.e.\ $\rm \zeta_{crp}=2\times 10^{-16}\,s^{-1}$). There is some evidence that the ionization rate is depth dependent in clouds, varying from $\rm \sim 2\times 10^{-16} \,s^{-1}$  in low $A_V$ clouds to values of $\rm \sim 2\times 10^{-17} \,s^{-1}$ in the centers of molecular clouds \citep[e.g.][]{2009A&A...501..619P, 2012ApJ...754..105H, 2012ApJ...745...91I, 2012A&A...537A...7R}.   As discussed in Paper I, the model has a high cosmic-ray ionization rate but equally good fits to the data can be obtained with the low cosmic ray rate.  However, we find that the higher cosmic ray rates produce a slightly higher \ci\ line intensity since the higher rates allow neutral C to extend deeper into the cloud. 

In addition to providing results from these models for the three principal lines observed in the data presented here (i.e.\ \coa\ \& \cob\ 1--0 and \ci\ 2--1) we also include the results for \ci\ 1--0 and \cii\ lines emitted at 492\,GHz and 1.9\,THz, respectively.  This is so that they may also be applied towards interpreting measurements made with sub-mm telescopes in Chile such as Nanten2, APEX and ALMA \citep[e.g.][]{2008stt..conf..488G, 2014ApJ...797L..17L}, as well as THz frequency telescopes.  These latter include airborne \citep[(e.g.\ SOFIA;][]{2012A&A...542L..13P} and space \citep[e.g.\ Herschel;][]{2010A&A...521L..18V, 2014A&A...561A.122L} facilities as well as future telescopes under development for Antarctica such as STO--2 \citep{2010SPIE.7733E..0NW}  and DATE5 \citep{2013RAA....13.1493Y}.

\subsection{PDR Model Comparison}
\label{sec:pdrmodelplots}
We present here the results for a representative model using our `standard' parameter set with $A_{V,\rm tot}=6$ mags. The abundance of  C$^+$, C, \coa\ and \cob, as well as of H$_2$, as a function of the optical extinction, $A_V$, from the cloud surface is shown in Fig.~\ref{fig:xh2covavplt}.  Both the ionized and neutral carbon are seen to be confined to the surface of the PDR, with the C$^+$ peaking at the front surface, and mostly disappearing by $A_V \sim 0.6$~mags, where the C peaks.  Further in to the cloud the carbon is converted into CO, with \cob\ rising rapidly between $A_V\sim 0.5 - 1$~mags.\ followed by a slow rise to $A_V \sim 1.8$~mags.  At this point photo-desorption of CO from the cold grains diminishes due to dust attenuation of the FUV and freeze-out of the CO onto grain mantles begins, with the CO abundance dropping by a factor of $\sim 3$ at $A_V \sim 3$~mags.\ (at the cloud center).  The  hydrogen gas, on the other hand, is converted almost entirely into H$_2$ by $A_V = 0.1$~mags.\ so the C is mostly found in regions where the hydrogen is molecular H$_2$ rather than atomic H\@.

The optical depth of the \ci, \cii\ and CO lines as function of $A_V$ in this representative model is shown in Fig.~\ref{fig:tauci13covavplt}.  The total extinction through the model cloud is $A_{V,{\rm tot}} = 6$~mags., and lines here are plotted to cloud center. Thus, the total optical depth in the sightline through the cloud is twice that shown at $A_V = 3$ in the Figure. \cii\ rises rapidly until $A_V \sim 0.5$ and then remains constant, as by then all the C$^+$ has been converted to C or CO\@.  The \ci\  lines rise rapidly to $A_V\sim 0.8$ past the abundance peak, and then only gradually thereafter as the \ci\ abundance falls. The \cob\ continues to rise within the cloud but starts to flatten near $A_V\sim 2.5$ due to \cob\ freeze-out. All the associated lines for these three species remain optically thin to cloud center. Through the full cloud extent (i.e.\ $A_V = 6$) the \cii\ and \ci\ 2--1 lines remain thin, while the \ci\ 1--0 and  \cob\ lines are marginally thick.  \coa\ is strongly optically thick beyond $A_V \sim 0.8$.

The gas temperature, $T_{\rm gas}$, and dust temperature, $T_{\rm dust}$, through the PDR are shown as a function of the 
extinction for the representative model in Fig.~\ref{fig:tgasplt2}.  The gas temperature averages around 30\,K within the first magnitude of the extinction, though rises slightly to $\sim 37$\,K for $A_V < 0.1$, at the very front surface of the cloud. 
The temperature rise is due to molecular hydrogen formation. The collisional rate coefficient of molecular hydrogen exciting C$^+$ is about half that of atomic hydrogen exciting C$^+$\footnote{We note that the collision rates we use for C$^+$ with H$_2$ are within a few percent of the recent rates of para--H$_2$ with C$^+$ in \citet{2014ApJ...780..183W}.  Adopting these rates results in only a few percent decrease in the gas temperature.  We also note that our collision rate of H with C$^+$ \citep[from][]{1990mcim.book.....F} is slightly higher than that in \citet{2005ApJ...620..537B}.  Adopting this more recent rate would further increase the gas temperature in the atomic H layer but since collisions with atomic hydrogen dominate in only a narrow region at the outer edge of the cloud this makes a negligible difference to our results.}.  Since the C$^+$ is the dominant coolant, the lower excitation rate leads to warmer gas.
At deeper layers, the temperature falls due to extinction of  FUV radiation field resulting in a lower photoelectric heating rate. 
At $A_V > 2$, the photodesorption of molecules from grain surfaces falls sufficiently so that CO stays frozen onto grains.
The freeze-out of the dominant coolants cause the temperature to rise slightly at cloud center.

The integrated line intensities for the \ci, \cii\ and CO lines through the PDR as a function of column density  (i.e.\ $A_V$) for the representative model are shown in Fig.~\ref{fig:ci13coTkmvAv}.  The intensity plotted is the total intensity emitted through one side of the cloud, including the emission from both the near and far sides (i.e., the emission generated at the far side of the cloud that passes through the cloud center and emerges at the near side of the cloud). These fluxes increase with depth into the PDR, until the carbon has been converted into CO, and the CO then frozen-out onto grains.  The model line fluxes for the \ci\ 2--1, \coa\ and \cob\ lines are typically 6--12 times those measured in the individual voxels.  However, the model uses a FWHM of 4 km/s, a factor 6 greater than the voxel width of 0.7 km/s used in the data analysis, and so is consistent with flux filling factors in the 2 arcmin beam of between $\sim 50$ and 100\%.

In Fig.~\ref{fig:ci13covAv3} are shown plots of the \ci\ 2--1 / \cob\ line ratio as a function of optical extinction, $A_{V,{\rm tot}}$, through molecular clouds for a variety of values of the input parameters. For each curve we keep the parameters fixed for the `standard' model, and vary only the parameter that is listed in the figure legend. The $\rm PAH/H$ curve is for a PAH abundance that is twice as high as for the standard model while the $\rm S/H$ curve is for a sulfur abundance that is twice the abundance used in our previous modelling \citep{1999ApJ...527..795K} and eight times larger than in our standard case. 
The $G_0$ curve is for a radiation field that is the local Galactic field, and the thermal pressure curve is a factor of 3 lower than the standard model. The parameters range roughly between those expected for nearby clouds and those in the inner Galaxy.  The higher PAH/H and S/H curves increase the \ci/\cob\ ratio by increasing the C abundance as described in \S\ref{sec:pdr}. The lower $G_0$ model produces a lower \ci/\cob\ ratio due to a lower heating rate in the \ci\ region. For the low pressure models, the lower density pushes the molecular transition deeper into the cloud, thereby reducing the \cob\ column for the low $A_{V, \rm{tot}}$ models.

The line ratio is seen to decrease as the cloud column density increases, since the \ci\ emission is confined to the front surface, $A_V < 0.6$~mags.\ (see Fig.~\ref{fig:ci13coTkmvAv}), whereas the \cob\ emission extends deeper into the cloud. Except for the low thermal pressure model,  beyond $A_{V{\rm tot}} \sim 4$~mags.\ the line ratio remains constant, indicating that neither line is contributing significant additional line flux for thicker clouds. All models obtain the observed \ci/\cob\ ratios of $< 2$ for sufficiently thick clouds. The lowest allowed column density cloud is $A_{V,{\rm tot}}\sim 1$, for the low FUV field model.
  
The dependence of the \coa/\cob\ line ratio as a function of the total optical depth through the cloud, $A_{V,{\rm tot}}$ is shown in Fig.~\ref{fig:co12co13vAv}.  Again, the line ratio is high for low optical depths, but rapidly decreases as $A_V$ rises, plateauing beyond $\sim 4$~mags., similarly to \ci/\cob.  In this case, however, this is caused by the freeze-out of the CO molecules onto dust grains. 

The dependence of these two line ratios (\coa/\cob\ and \ci\ 2--1/\cob) are also shown as a function of the \cob\ line flux in Figs.~\ref{fig:co12co13vco13} and \ref{fig:cico13vco13}. Similar behaviour to the plots against extinction, $A_V$, are exhibited, with the line ratios being high for small values of the \cob\ line flux.  This reflects, correspondingly, the CO optical depth being lower for thin clouds (i.e.\ $A_v < 2$), and the \ci\ emission arising from the same region, near to the cloud surface. In practice, however, observing such high ratio but low intensity gas in our data set will be difficult since one or both lines may not be detectable.    We also caution that low intensity gas could have low ratios, if the beam filling factor is small.  In principle, however, it is possible to have gas with \coa/\cob\ and \ci/\cob\ ratios much higher than those observed in the G328 data set presented here.  They would also be more readily detectable with higher angular resolution measurements, able to separate such regions out from high column density sightlines.

We also calculated the `dark gas' contribution in the standard model. Using the definitions in \cite{2010ApJ...716.1191W}, we assume the dark gas layer begins when the gas is half molecular (i.e., $2n({\rm H_2})=0.5n$) and ends when the optical depth in the ${\rm CO\, (1-0)}$ transition equals one. We find the line-of-sight dark gas fraction is 27\% of the total column density of this $A_{V,tot} = 6$\,mags.\ slab. Since the ${\rm H_2}$ forms essentially at the cloud surface ($A_V \sim 0.1$), the dark gas fraction of the total molecular column density is also about 27\%.  We can also estimate the fraction of dark gas in a spherical cloud by converting the $A_V$ steps to radius steps ($\delta r = \delta A_V(r) \, 2.0 \times 10^{21}/n(r)$, where $\delta r$ and $\delta A_V$ are the distances between radius and $A_V$ grid points, respectively). The mass at radius $r = \Sigma \delta r$ is found by  summing the density in spherical shells $M(r)= \Sigma n(r)4\pi r^2 \delta r$ (H ${\rm cm^{-3}}$). We find a fraction of dark gas compared to the total molecular gas is $\sim 67$\% with a similar fraction when the total (molecular plus atomic) mass is included. The dark gas fraction is a factor of two higher than that found by \cite{2010ApJ...716.1191W} for local Galactic GMCs. 
However, the mean area-averaged column density for the model cloud is found to be $\bar{A_V} = 2.7$. This is smaller than that for GMCs \citep[${\bar A_V}\sim 8$;][]{1987ApJ...319..730S} and the dark gas fraction is predicted to increase as the mean column density decreases \citep{2010ApJ...716.1191W}.

\begin{figure}
\includegraphics[angle=0,scale=0.7]{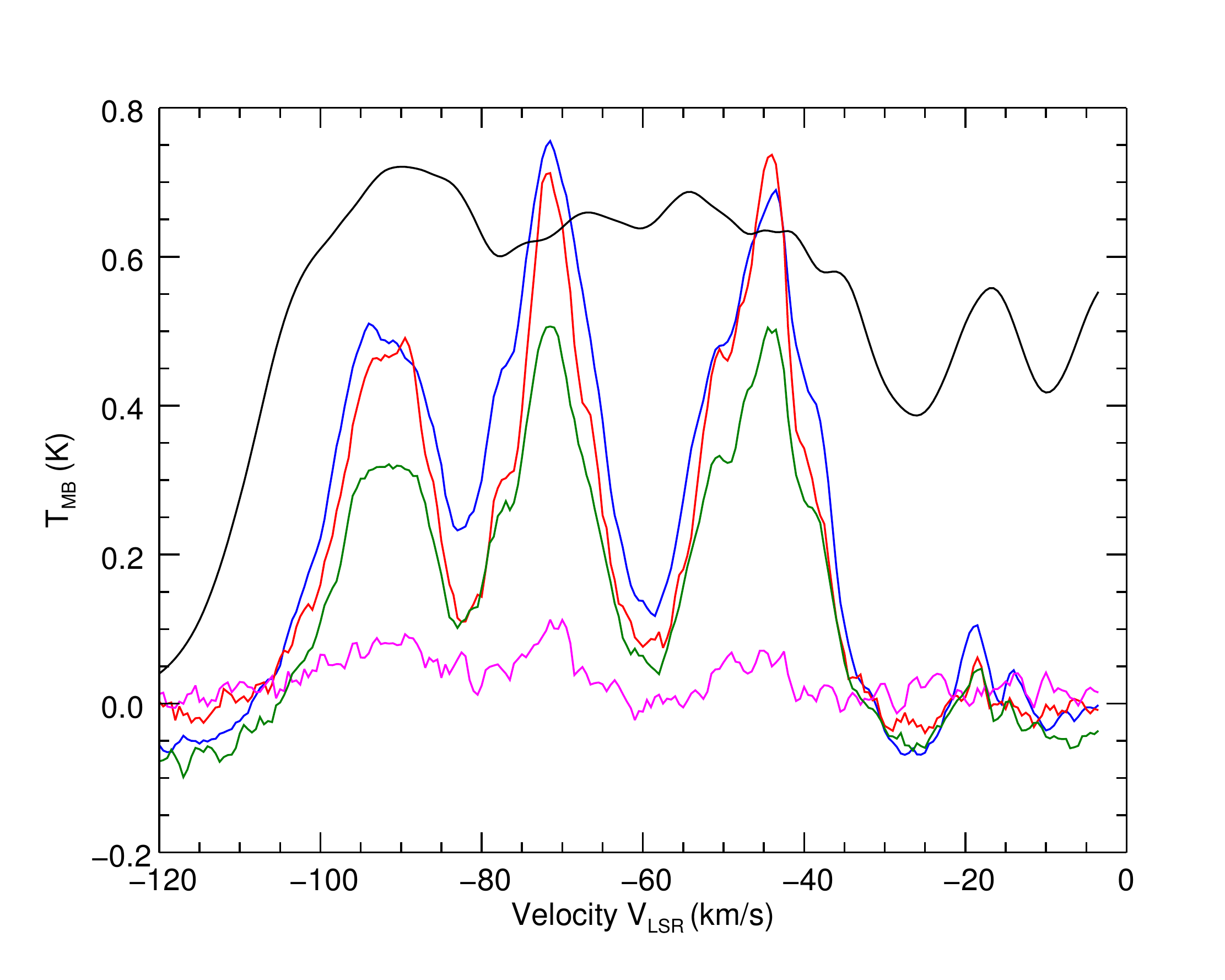}
\caption{Line profiles of the mean integrated emission from the entire region covered by the $\sim 1^{\circ} \times 1^{\circ}$ region of the HEAT G328 data cube. Lines are \coa/5 (blue), \cob\ (red), \coc\ (magenta), \ci\ (green) and HI/150 (black). Three principal spiral arm crossings can be seen in the CO and [CI] data between $-100$ to $-85$~km/s (Norma; near-portion of arm), $-80$ to $-65$~km/s (Norma; far-portion of arm) and $-55$ to $-40$~km/s (Scutum--Crux; near-portion), respectively. The weaker feature at $-20$~km/s corresponds to the far-portion of the Scutum-Crux arm. \label{fig:profiles}}
\end{figure}

\begin{figure}
\hspace*{1.5cm}
\includegraphics[angle=0,scale=0.6]{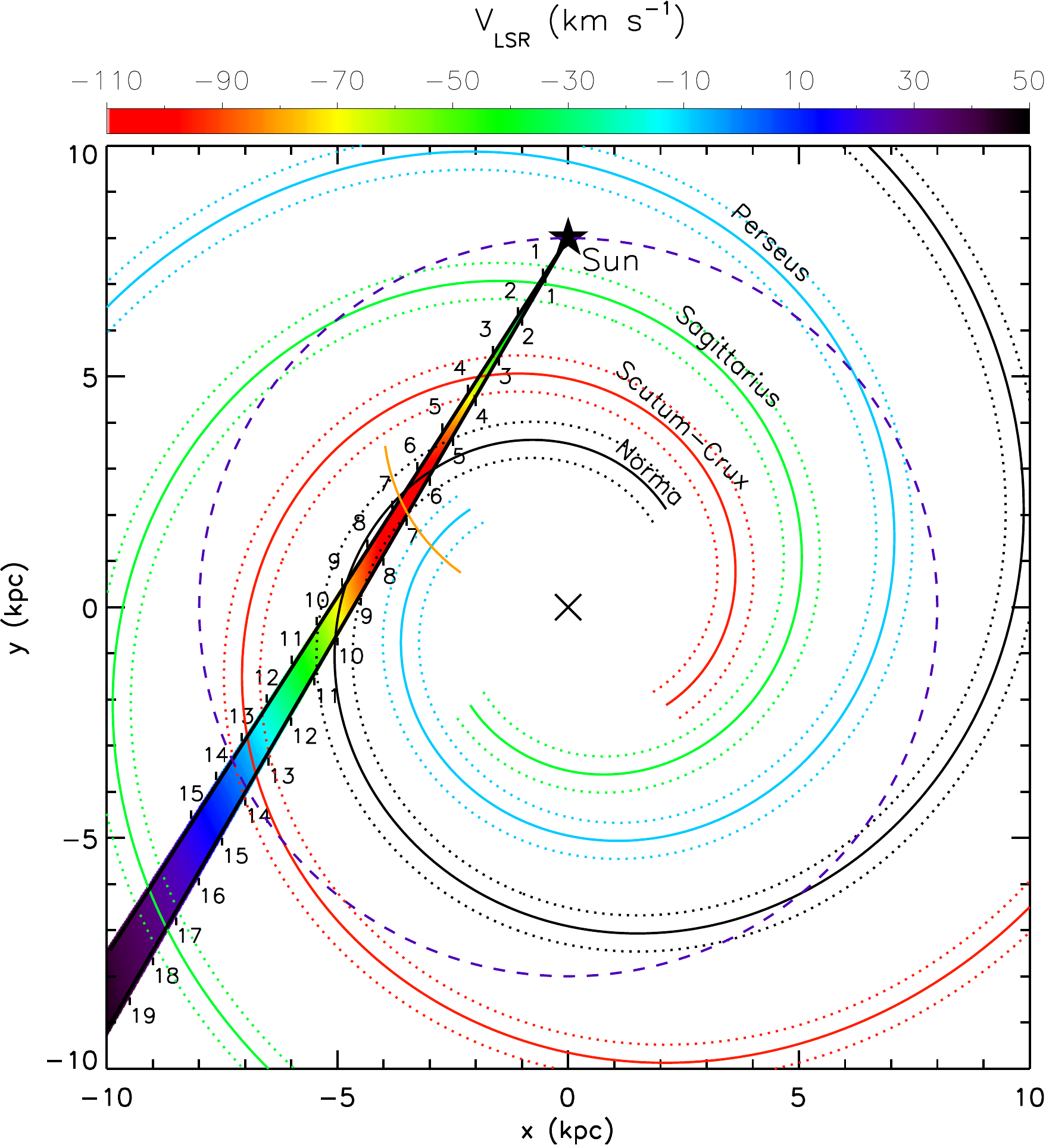}
\caption{Schematic of a four-arm model of the Galaxy to aid in visualization of the data along the G328 sight line.  The model uses the parameters from \citet{2014AJ....148....5V} with the spiral arms shown by color lines:  Perseus (turquoise), Sagittarius (green),  Scutum-Crux (red) and Norma (black).  Their extent is indicated by the corresponding dashed lines.  The G328 sightline from the Sun is indicated by the wedge, with the Solar Circle (at $R_{\odot} = 8$\,kpc) shown by the dashed purple line (the cross indicates the Galactic center).  The color shading within the wedge shows the expected line of sight radial velocities ($V_{\rm LSR}$) using the galactic rotation model of \citet{2007ApJ...671..427M} for the Fourth Quadrant, matched to the \citet{1993A&A...275...67B} model for the outer Galaxy (i.e.\ positive velocities, when $R > R_{\odot}$).  The short orange arc indicates the locus for the tangent point (where radial velocities are at their most negative).  The spatial scale along the axes is in kpc, and the numbers along the wedge show the distance from the Sun, also in kpc.    \label{fig:galaxymodel}}
\end{figure}

\begin{figure}
\includegraphics[angle=0,scale=0.18]{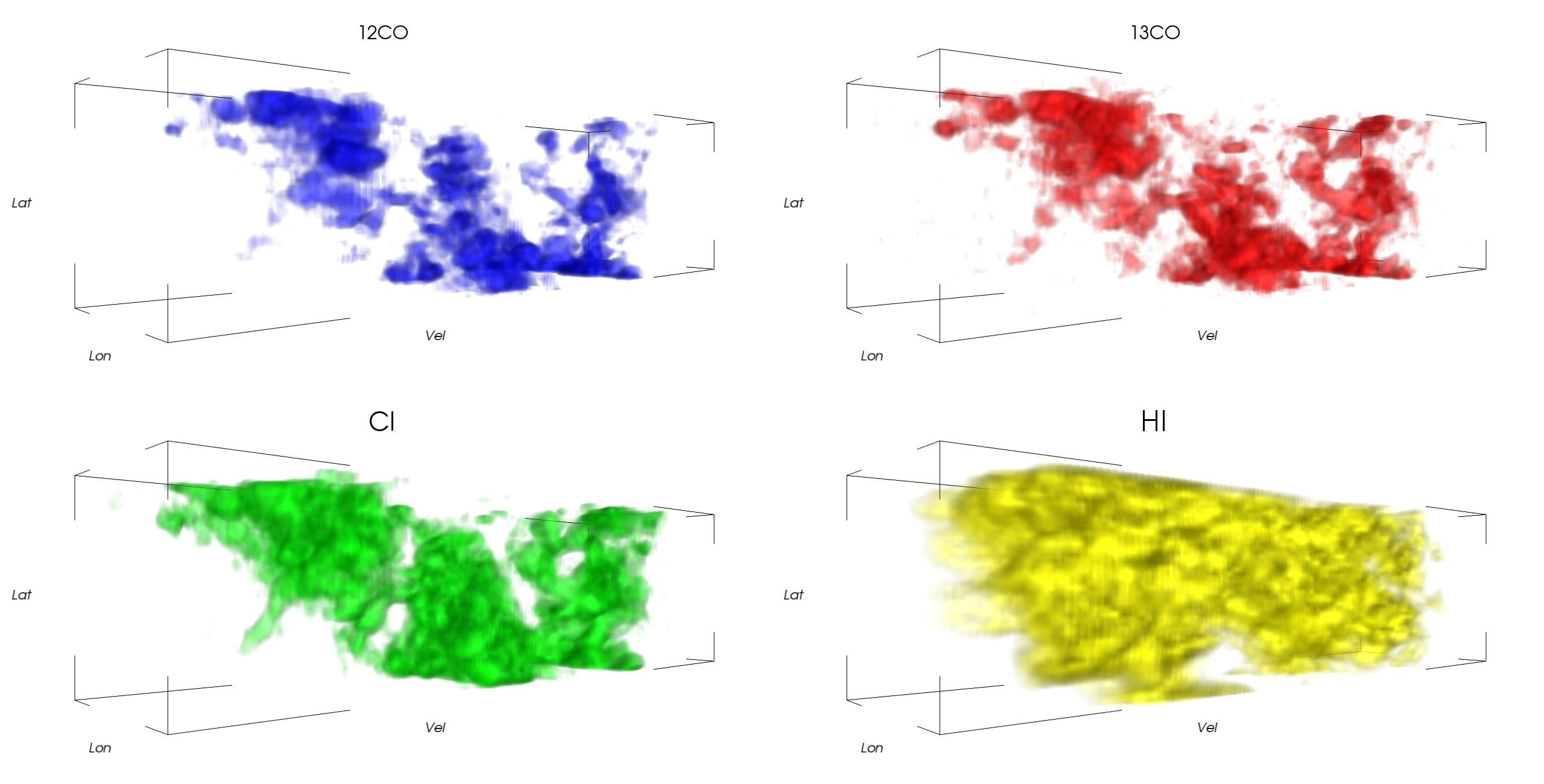}
\caption{Renderings of the G328 data cubes for \coa\ (blue), \cob\ (red), \ci\ (green) and HI (yellow).  Galactic longitude and latitude are along the short axes, with velocity along the long axis (with the most negative velocities to left). The three principal arm crossings along the sight line can be distinguished: Norma near-, Norma far- and Scutum-Crux near-, going from left to right.  These are also shown as moment images in Fig.~\ref{fig:momentimages120}.  The relative extent of the atomic and molecular gas can be gauged, with the atomic hydrogen enveloping both the carbon and carbon monoxide emitting-gas.  \label{fig:3dview}}
\end{figure}

\begin{figure}
\vspace*{-1.5cm}
\includegraphics[angle=0,scale=0.25]{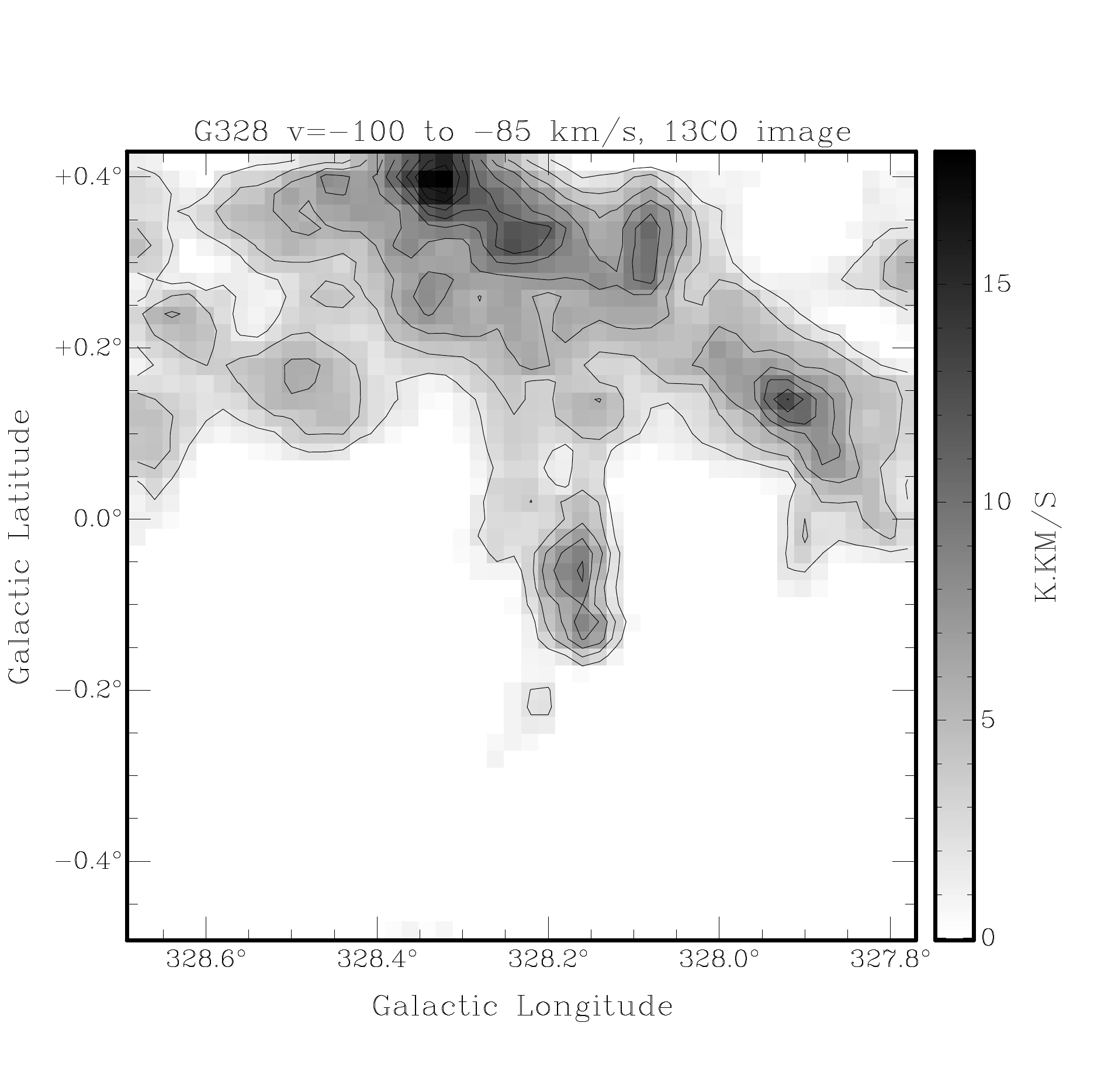}
\hspace*{-0.4cm}\includegraphics[angle=0,scale=0.25]{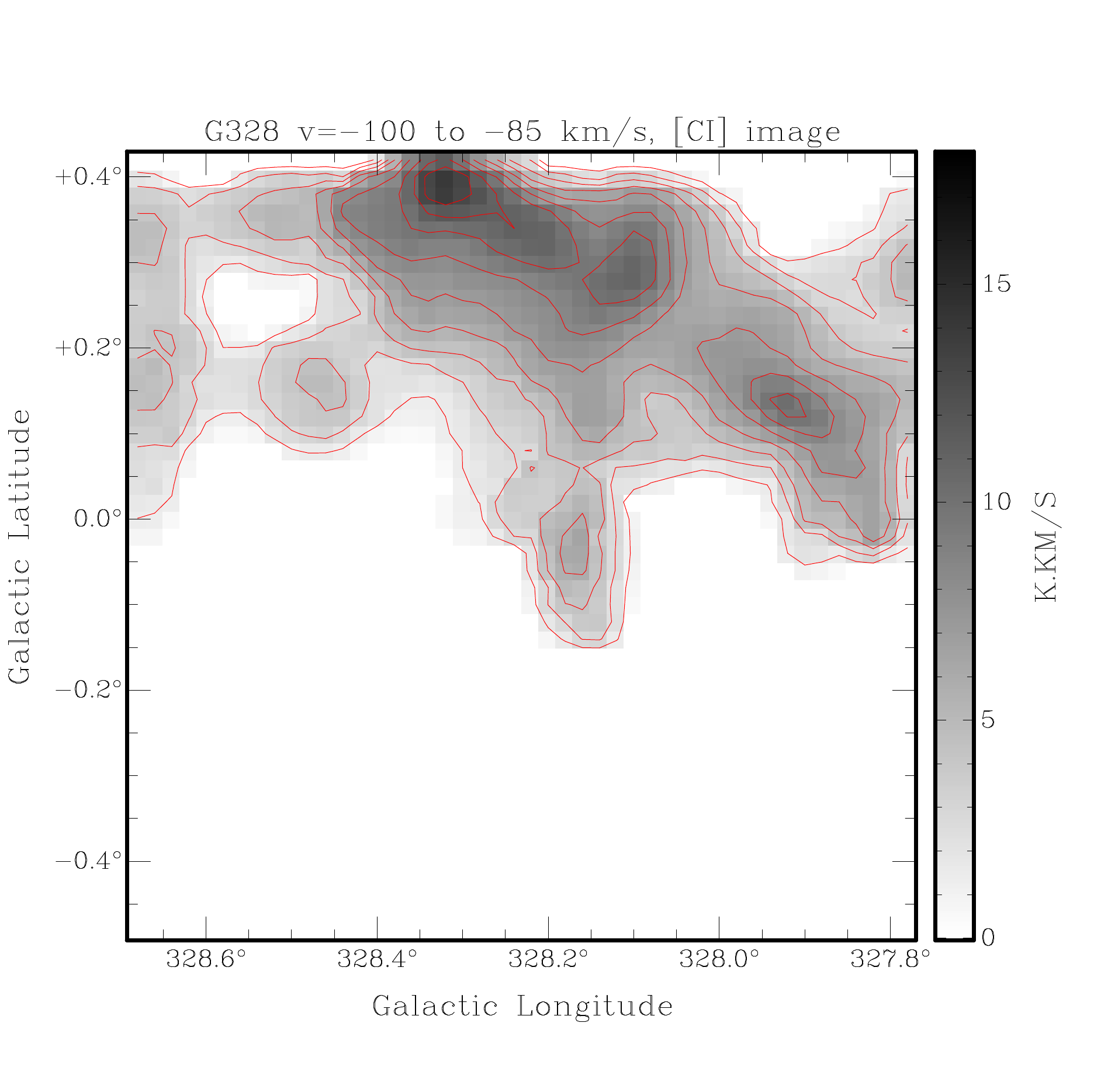}
\hspace*{-0.4cm}\includegraphics[angle=0,scale=0.25]{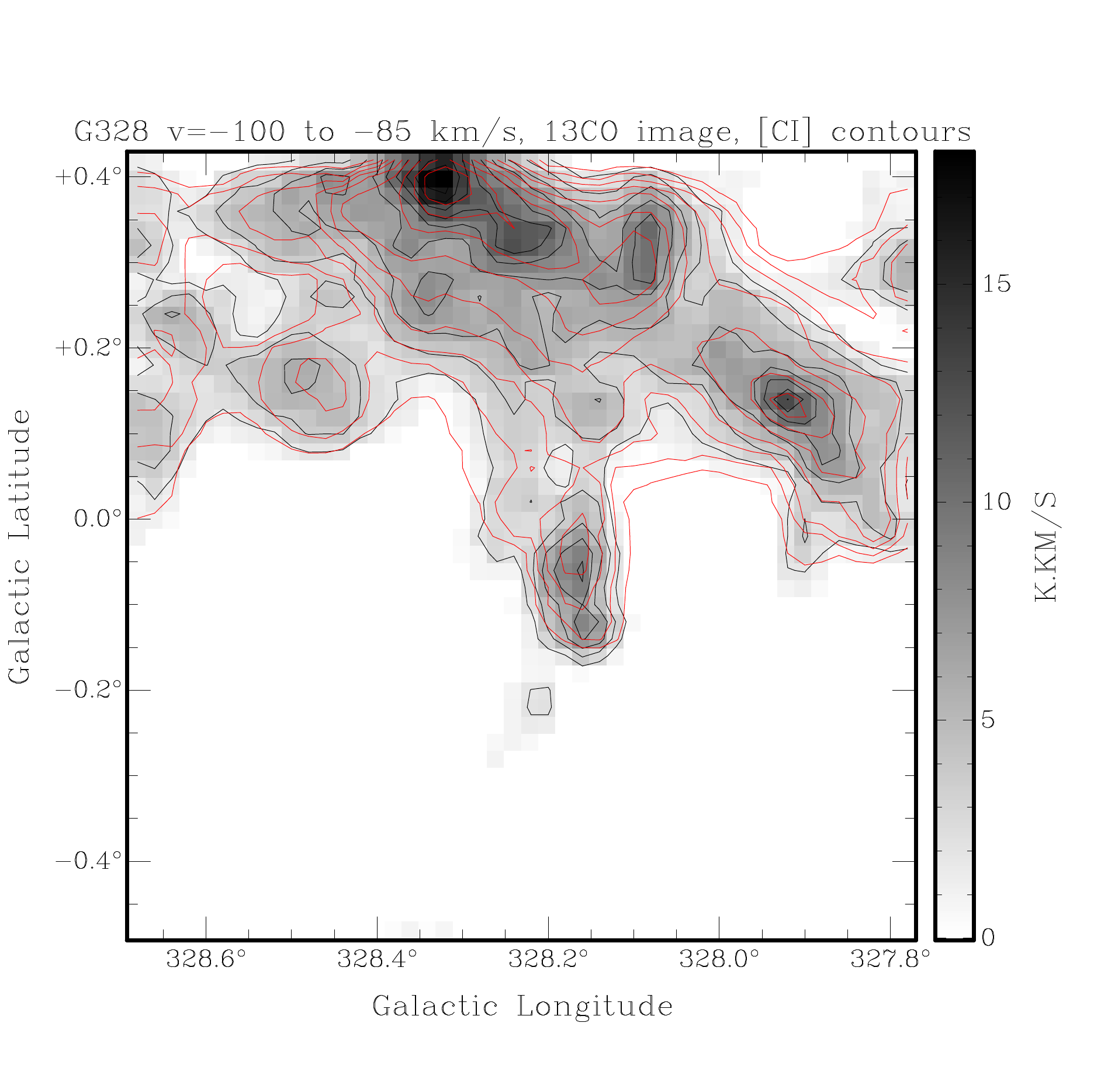}\\
\includegraphics[angle=0,scale=0.25]{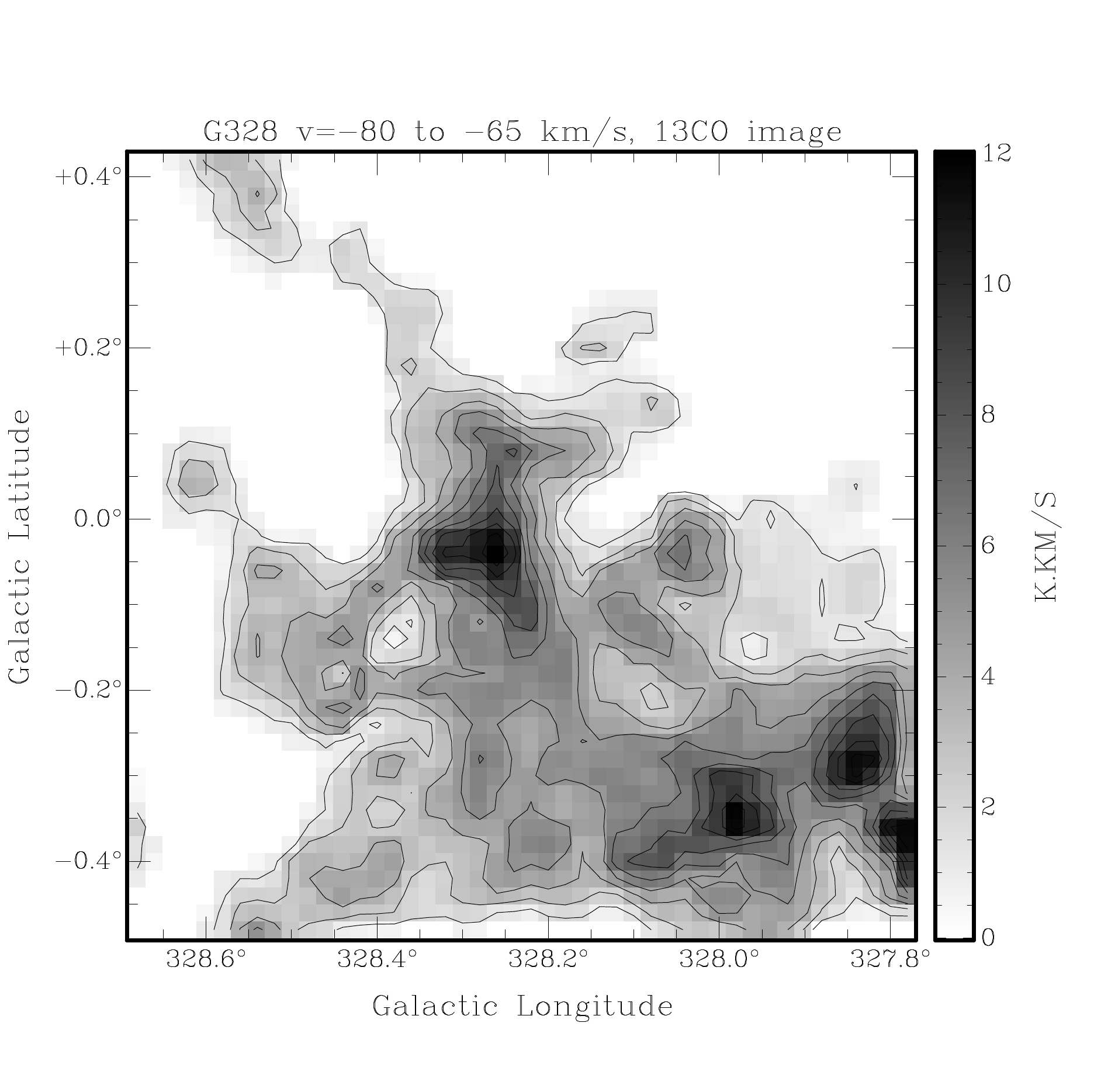}
\hspace*{-0.4cm}\includegraphics[angle=0,scale=0.25]{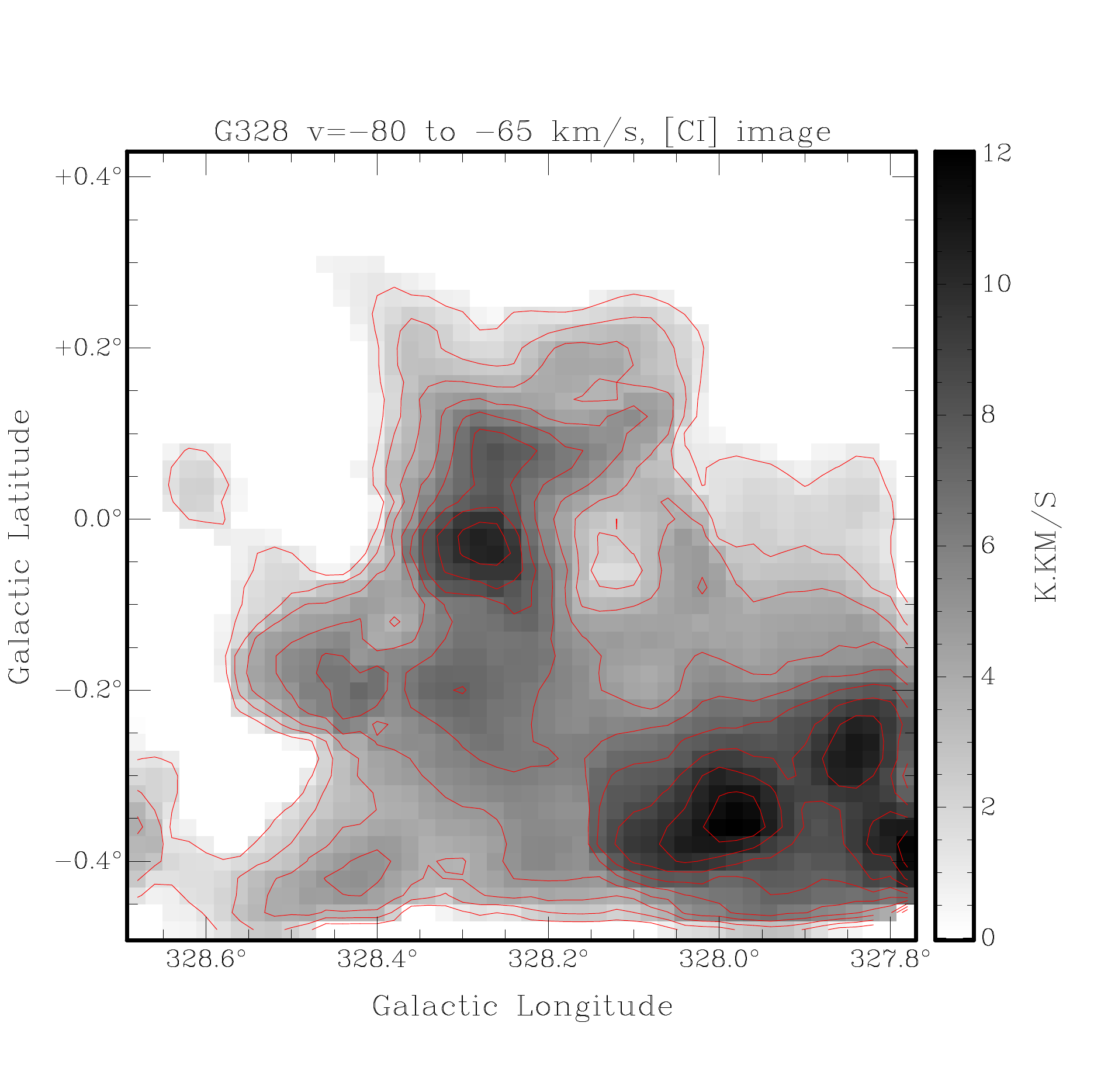}
\hspace*{-0.4cm}\includegraphics[angle=0,scale=0.25]{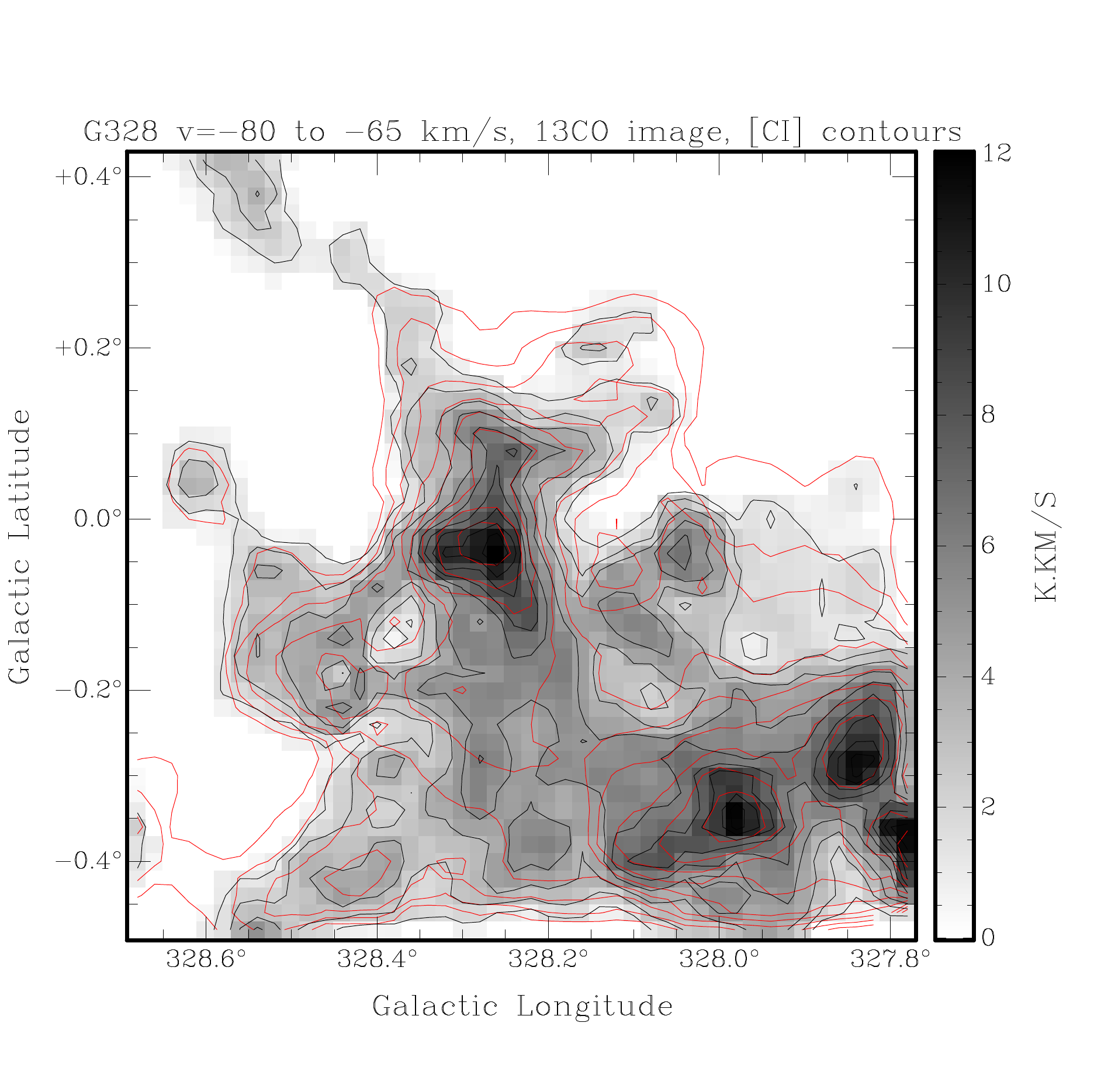}\\
\includegraphics[angle=0,scale=0.25]{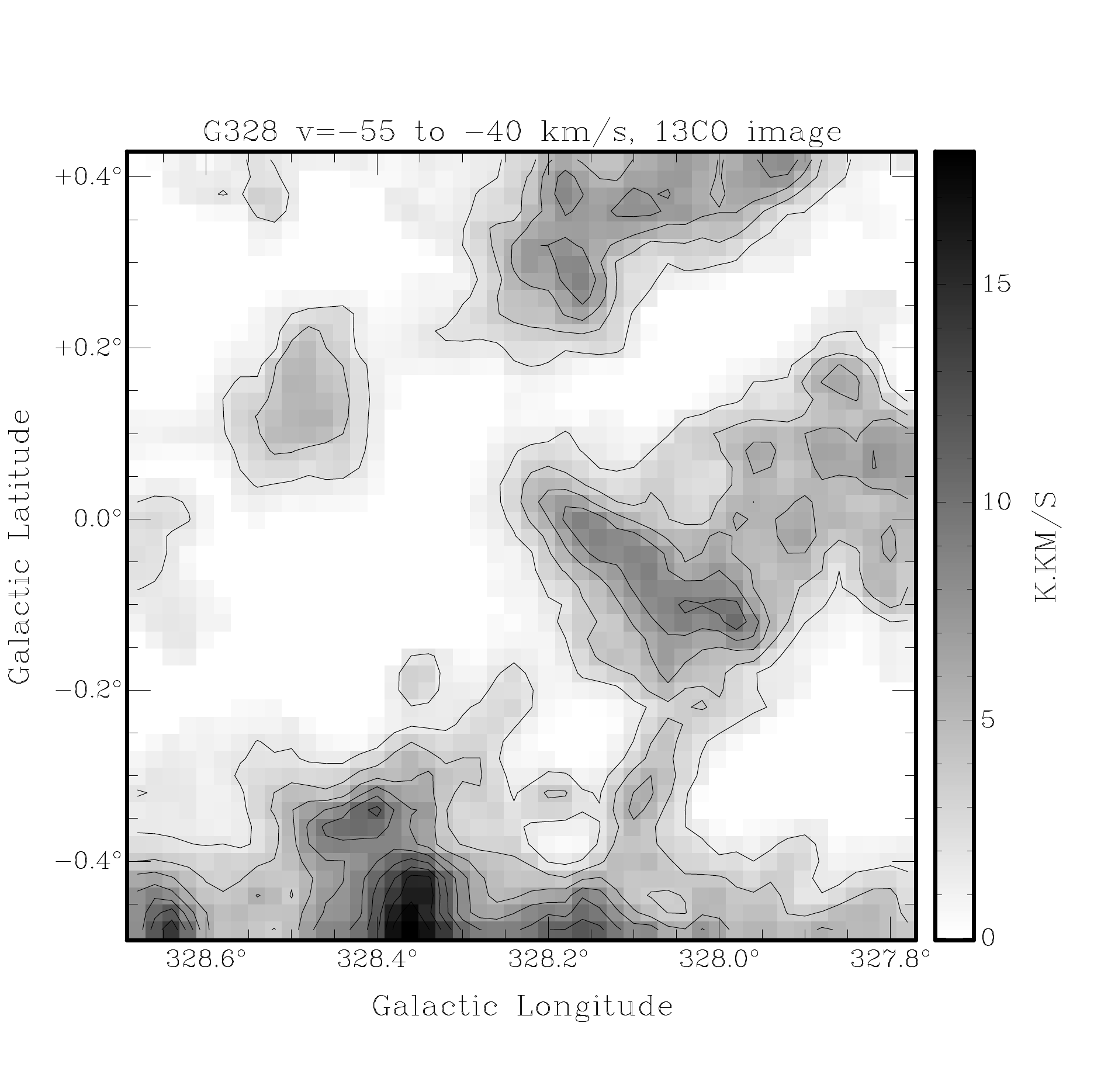}
\hspace*{-0.4cm}\includegraphics[angle=0,scale=0.25]{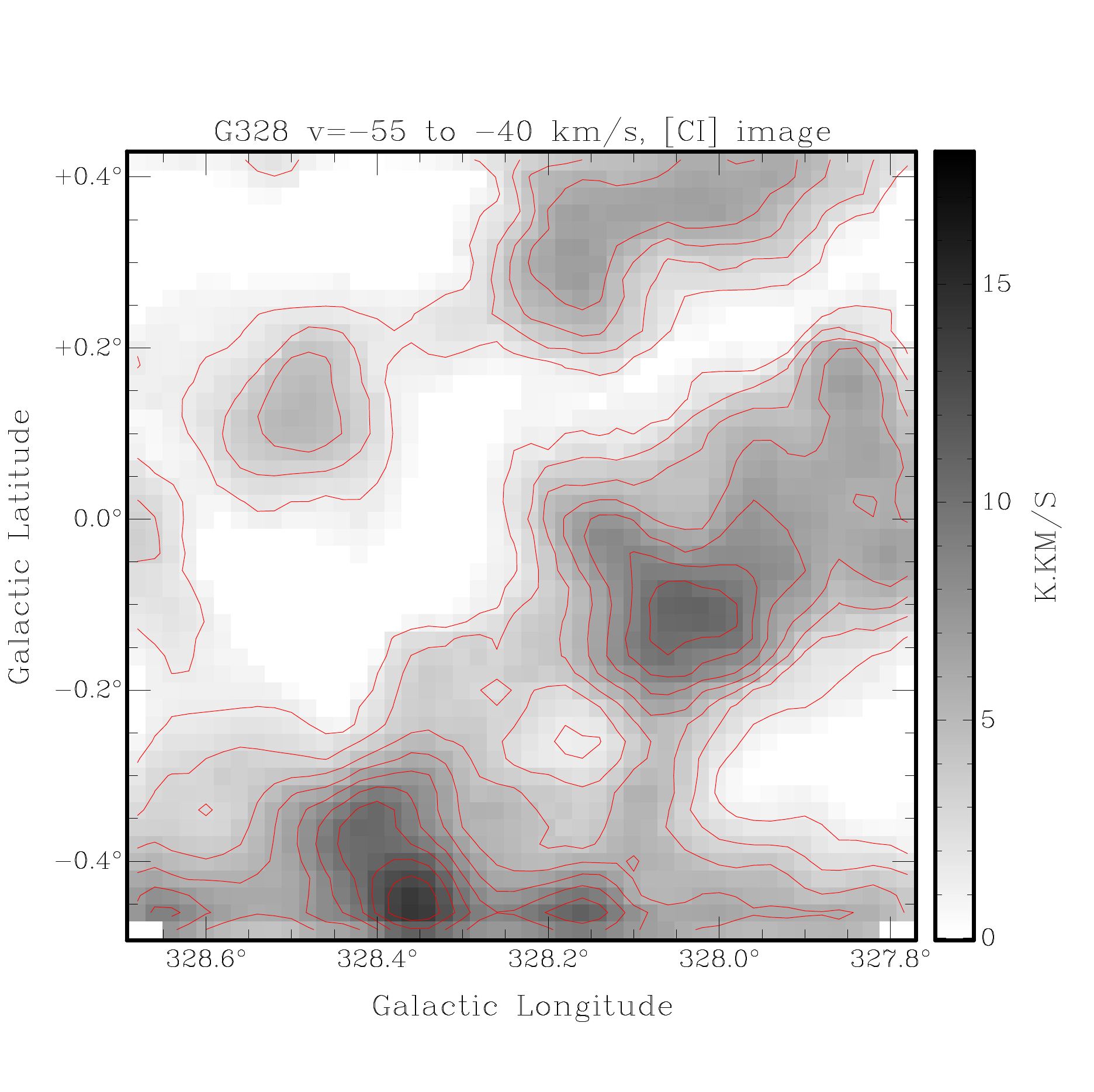}
\hspace*{-0.4cm}\includegraphics[angle=0,scale=0.25]{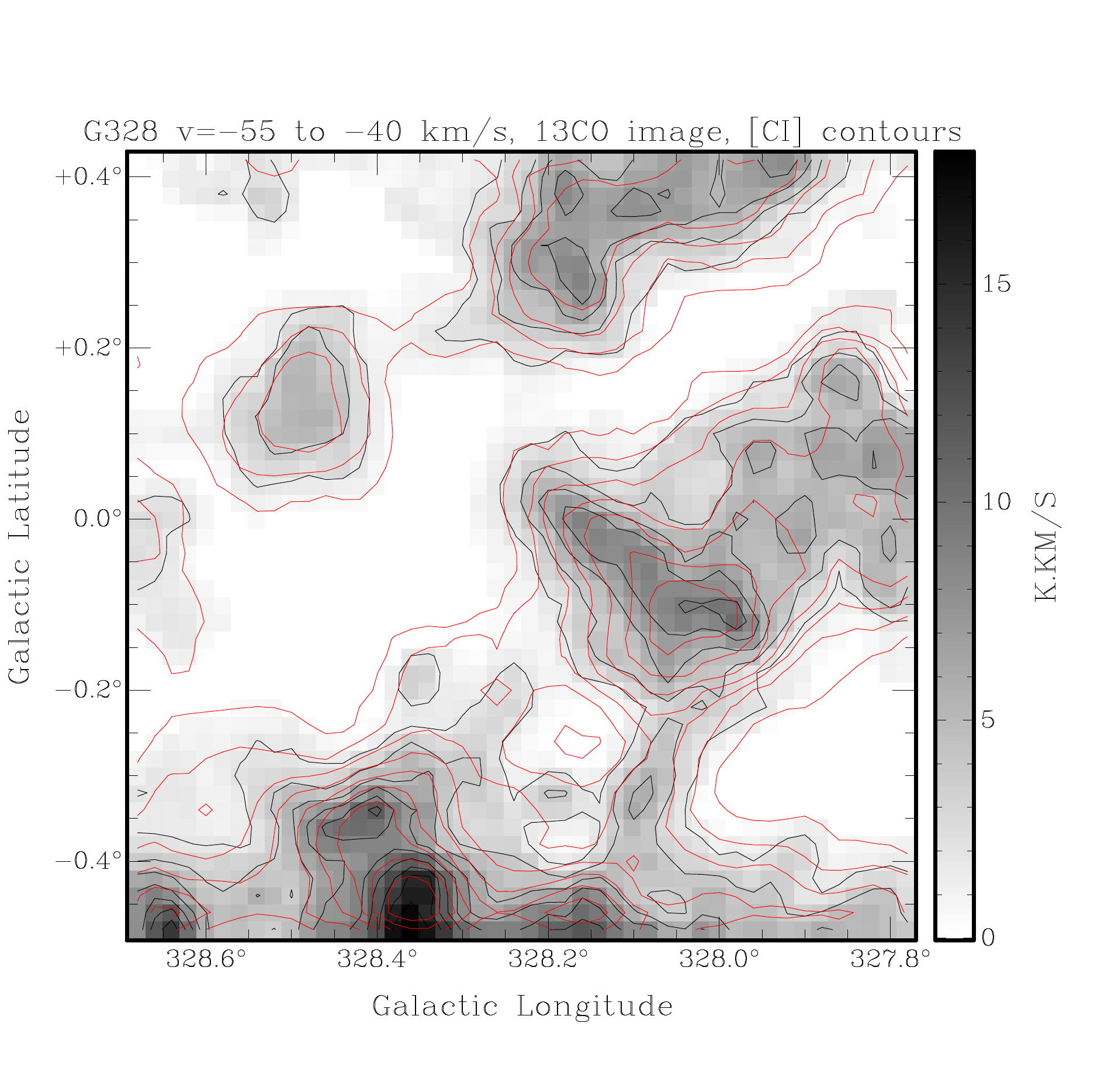}\\
\includegraphics[angle=0,scale=0.25]{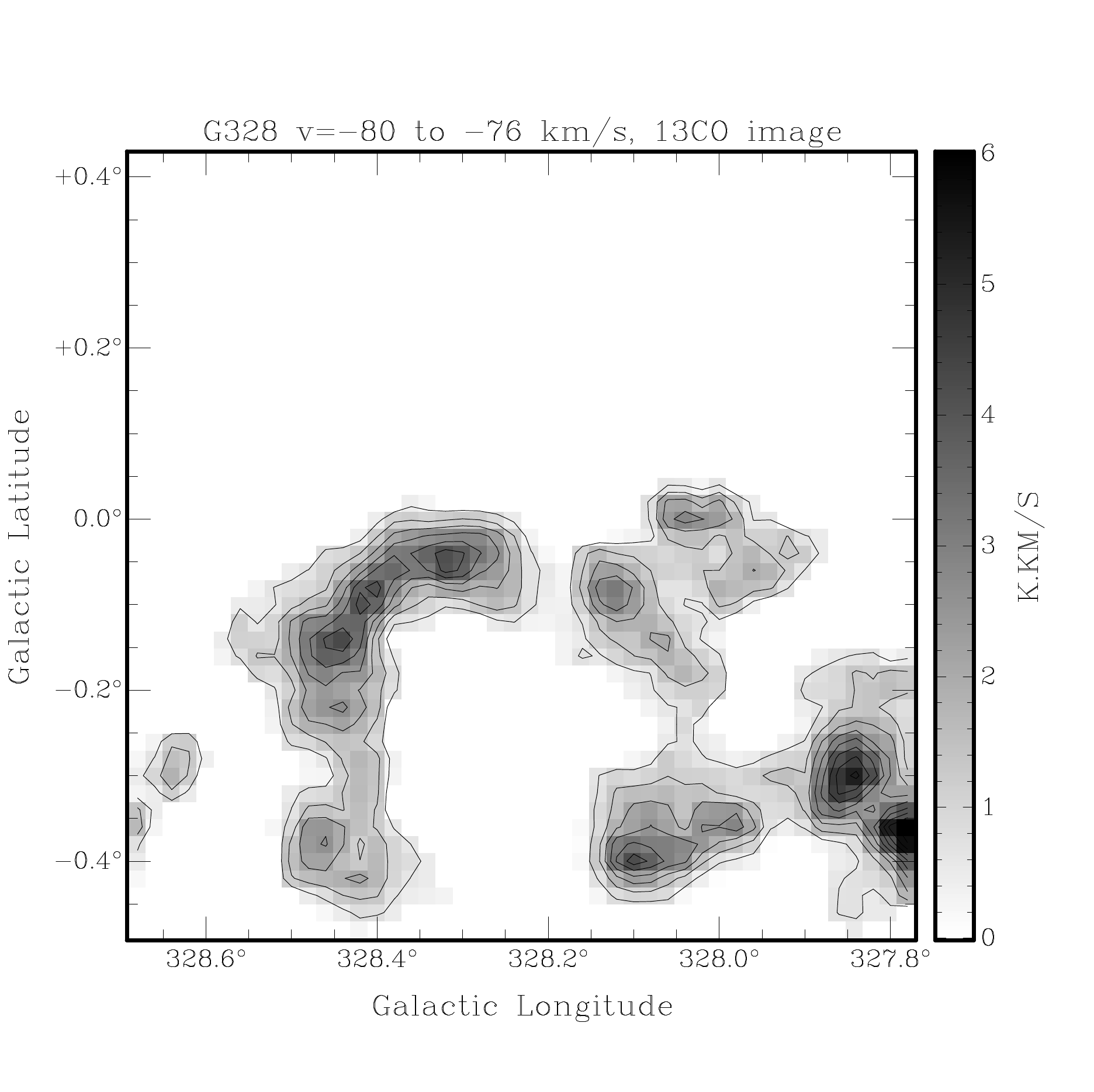}
\hspace*{-0.4cm}\includegraphics[angle=0,scale=0.25]{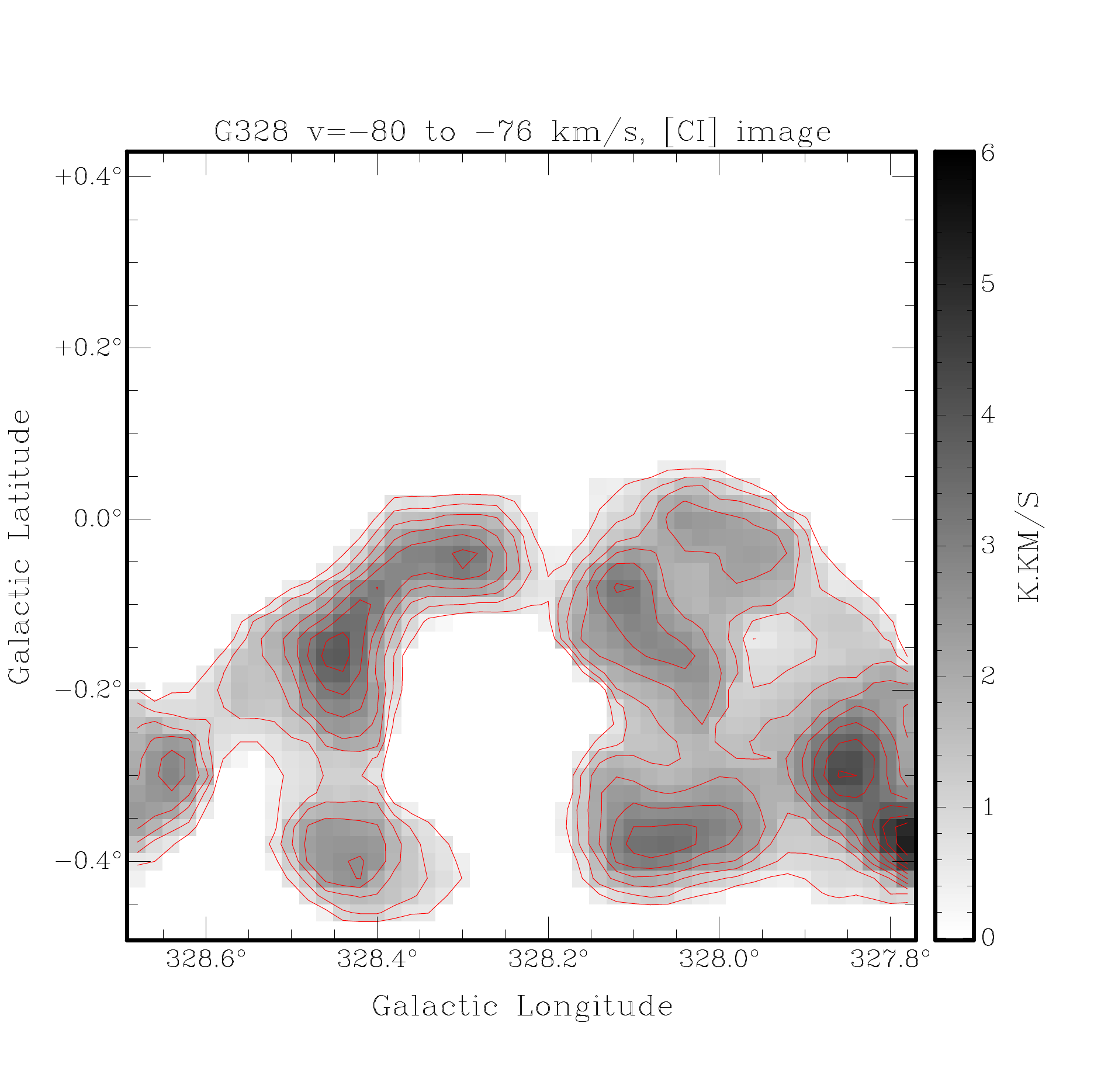}
\hspace*{-0.4cm}\includegraphics[angle=0,scale=0.25]{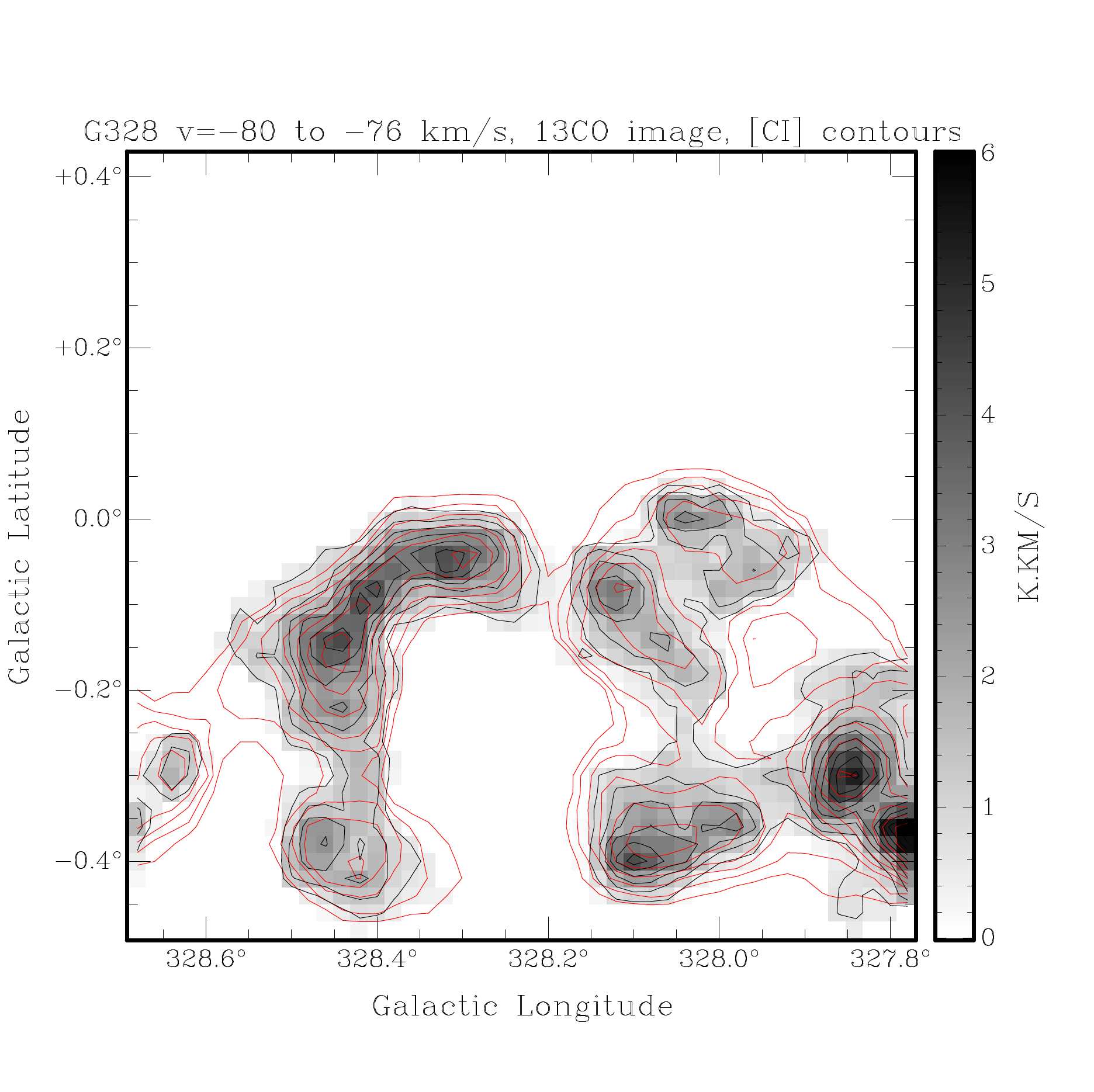}\\
\vspace*{-1.5cm}
\caption{Across: (i) Column 1: \cob\ moment 0 images (i.e.\ integrated flux), overlaid with their own contours (in black, at every 10\% of the peak value), (ii) Column 2: \ci\ moment 0 images, overlaid with their own contours (in red, at every 10\% of the peak value)  and (iii) Column 3: \cob\ images overlaid with red \ci\ contours.  Down: Velocity ranges for (a) $-100$ to $-85$ km/s (Norma near), (b) $-80$ to $-65$ km/s (Norma far), (c)  $-55$ to $-40$ km/s (Scutum-Crux near) and (d) $-80$ to $-76$ km/s (the filament studied in Paper I).  The scale in the color bar refers to the flux in image in K km/s. The original datacubes were all first smoothed with a $120''$ FWHM Gaussian. \label{fig:momentimages120}}
\end{figure}

\begin{figure}
\includegraphics[angle=0,scale=0.4]{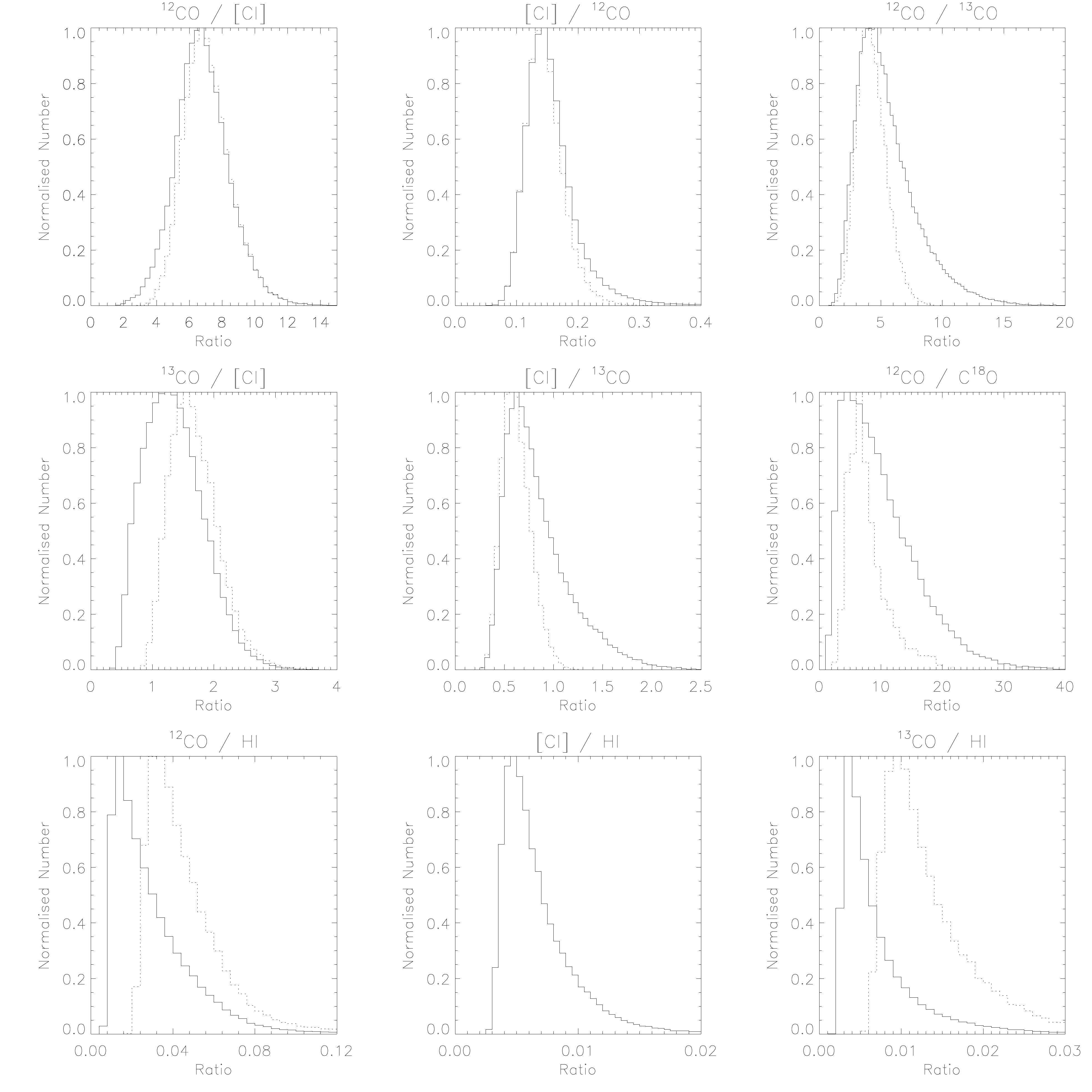}
\caption{Normalized histograms of the distributions of selected line ratios from the data set (\coa/\ci, \ci/\coa, \coa/\cob, \cob/\ci, \ci/\cob, \coa/\coc, \coa/HI, \ci/HI \& \ci/\cob, respectively, as labelled) determined on a voxel-by-voxel basis (i.e.\ pixel-velocity channel).  To be included the [CI] flux needed to be $> 3\sigma$ per voxel but the fluxes for each of the 3 CO lines (\coa, \cob, \coc)  only needed to be $> 1\sigma$ in the voxel (see Table~\ref{tab:voxels}). Overlaid as dotted lines are the corresponding distributions if a $3 \sigma$ threshold is applied for the three CO lines instead. Statistics relating to the distributions are listed in Table~\ref{tab:voxelstats}. \label{fig:histograms}}
\end{figure}

\begin{figure}
\includegraphics[angle=0,scale=0.25]{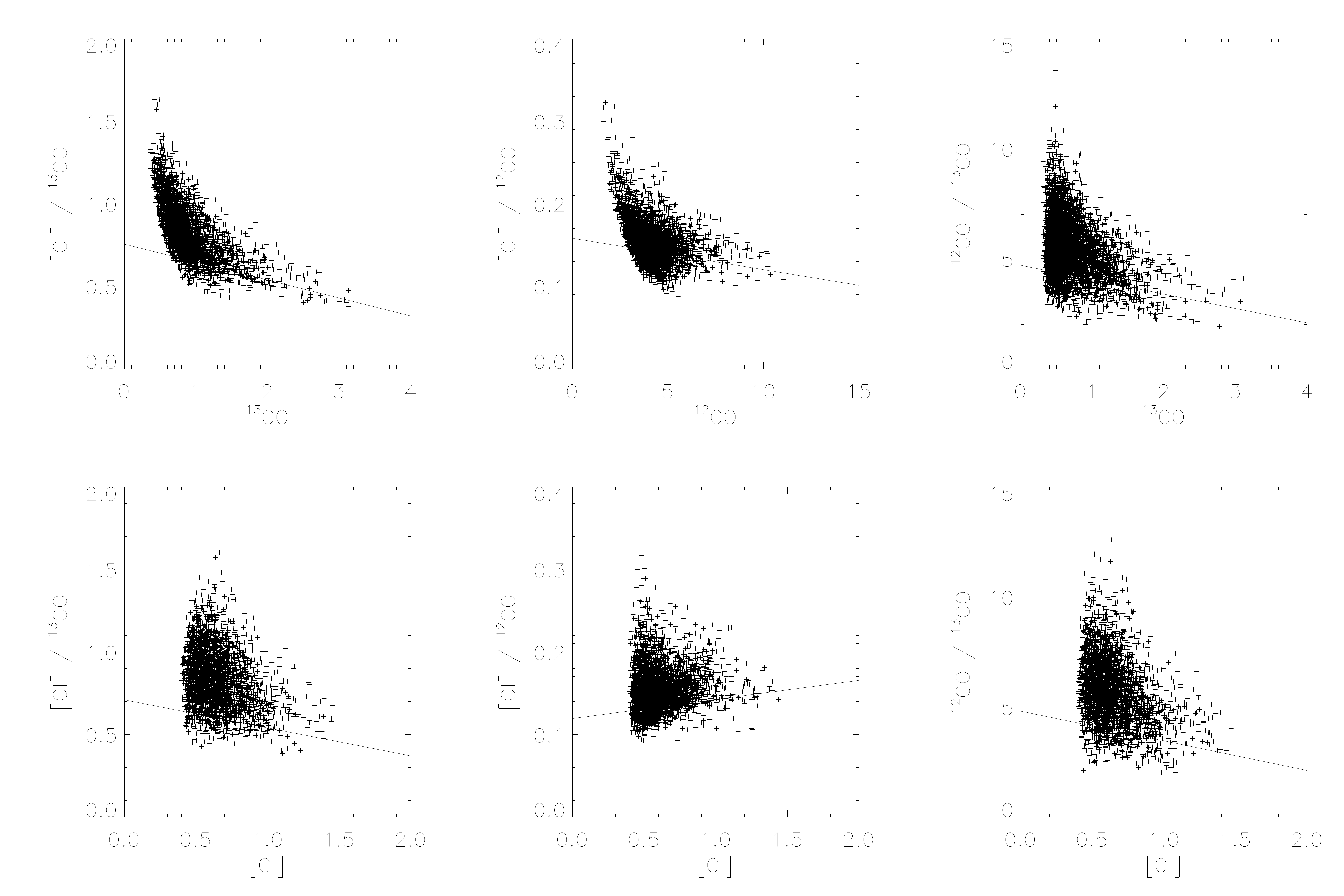}
\caption{Distributions of various line ratios vs. line intensities.  $1\sigma$ thresholds are applied to CO, $3\sigma$ for [CI] (see text).  Shown are, on top row, then second row: (a) [CI]/$^{13}$CO vs. $^{13}$CO, (b) [CI]/$^{12}$CO vs. $^{12}$CO, (c) $^{12}$CO/$^{13}$CO vs. $^{13}$CO, (d) [CI]/$^{13}$CO vs. [CI], (e) [CI]/$^{12}$CO vs. [CI] and (f) $^{12}$CO/$^{13}$CO vs. [CI].  Ten adjacent points have been averaged together in making these plots in order to  improve their clarity and error bars have also not been shown.  The lines show the best linear fit (weighted, minimum least squares) to the data; the coefficients (A, B), with {$Ratio = A \times Flux + B$} are as follows: 
$(0.8, -0.1), (0.2, -0.004), (5, -0.7), (0.7, -0.2), (0.1, 0.02)\, {\rm and}\, (5, -1.4)$, respectively [and, as discussed in the text, error bars are lower for the smallest ratios when the flux is low, hence the linear fits are weighted to pass through these points rather than the bulk of the points, which have lower S/N]. The number of points fitted in each plot is evident from Table~\ref{tab:voxelstats}. \label{fig:ratiovsflux}}
\end{figure}

\begin{figure}
\includegraphics[angle=0,scale=0.7]{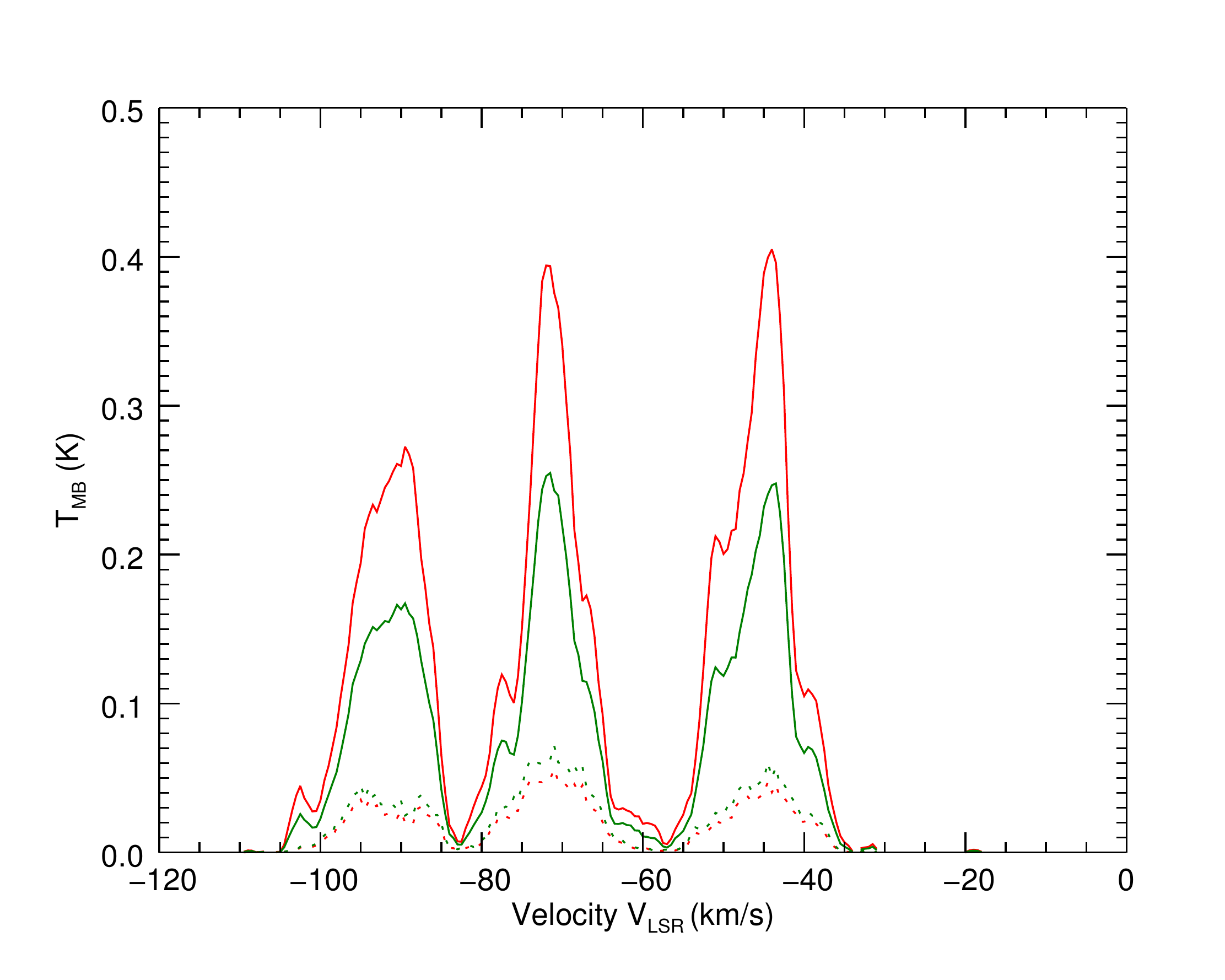}
\caption{Averaged line profiles for \cob\ and \ci\ over the entire data cube for all voxels that have ``high'' ($>1.0$) and ``normal'' ($<1.0$) \ci/\cob\ line ratios.   \cob\ is shown in red and \ci\ in green.  Voxels with ``normal'' ratio values are the solid lines and those with ``high'' ratios are dotted.   Clearly those regions with ``high'' ratios also have lower fluxes, on average. \label{fig:avprofiles}}
\end{figure}

Finally, we have estimated the optical depth of the (cold) atomic gas in the PDR through application of the radiative transfer equation $N = T_{S} \Delta V X (1 - e{^{-\tau}})$, where $N$ is the column density of atomic hydrogen in the PDR model (Table~\ref{tab:pdrmodeloutput}), $T_S$ its spin temperature (i.e.\ as at the front of the PDR; 35\,K from Fig.~\ref{fig:tgasplt2}), $\Delta V$ the FWHM line width (Table \ref{tab:pdrmodel}), and $X$ the HI X-factor ($\rm 1.8 \times 10^{18} \, cm^{-2} \, K^{-1} \,km^{-1} \,s$; \citet{1990ARA&A..28..215D}).  This yields $\tau \sim 0.3$ for the atomic gas in the PDR\@.  Following \citet{1999ApJ...517..209G}, the antenna temperature for the 21cm HI line emission from this gas is then given $N \ = T_{A} \, \Delta V X \frac{\tau}{1 - e{^{-\tau}}}$, which yields $T_A \sim 8$\,K\@.  This can be compared to the typical brightness temperatures observed along the G328 sightline, 90--100\,K (Fig.~\ref{fig:profiles}), which are dominated by hydrogen in (warmer) atomic gas clouds rather than in PDRs.  From their ratio we estimate that the fraction of atomic hydrogen found in PDRs along the sight line is $\sim 10\%$, with the remaining $\sim 90\%$ being in atomic clouds.  Furthermore, we may also estimate the amount of ``dark'' atomic gas in the PDR by calculating the column derived by applying the optically thin limit in the above formula, and comparing this to the actual column from the model.  We find that $\sim 15\%$ of the atomic gas in the PDR is dark; i.e.\ a column of H amounting to $\sim \rm 10^{19} \, cm^{-2}$.    This is also a negligible fraction of the total column of atomic gas along the G328 sightline, around $\sim 1\%$ of it.

\section{Discussion}
\label{sec:discuss}
The models presented above in \S\ref{sec:pdrmodelplots} show that, in the low--FUV field PDR environment typical of the diffuse molecular environment, that ionized C$^+$ is found within $A_V < 0.5$~mags.\ of the cloud surface. Neutral C extends as far as $A_V \sim 2$\,mags.\ into the PDR, but peaks in abundance at $A_V \sim 0.5$\,mags.  Molecular CO exists deeper in than $A_V \sim 0.5$\,mags. Freeze-out of CO begins by  $A_V \sim 2$\,mags.\, with only one-third of the molecule remaining in the gas-phase by $A_V \sim 3$\,mags.

Given the different extents where these species are found inside a PDR, the measured line ratios, such as \coa/\cob\ and \ci/\cob, will depend on the column density of the PDR sampled by the line of sight.  In particular, those sight lines that either only pass through their edges, or include low column density PDRs (i.e.\ only extending for 1--2~mags.\ rather than the more typical 8 mags.\ for GMCs) will display elevated values of these ratios.

Optical depth, however, also needs to be accounted for in determining line fluxes, and hence line ratios.  \ci, \cii\ and \cob\ remain optically thin within these emitting regions. However, the \coa\ line rapidly becomes optically thick further into the molecular gas from the position where it is first encountered.

Thus \cii\ line flux rise until a depth of $A_V \sim 0.5$~mags.\ is reached into a PDR (and then remains flat as the sightline penetrates further), while \ci\ rises until a depth of $A_V \sim 1$~mags.  In both cases a flattening in flux with further depth is because the species are not found further into cloud.  For the \coa\ line flux, in contrast, while it rises rapidly between $A_V = 0.5$ and 1 mags., it then saturates as it becomes optically thick.  Only modest rises in flux then occur going deeper.  On the other hand, \cob\ line fluxes continue to rise steadily with increasing depth into a PDR until CO freezes out.  Representative line fluxes and column densities, for a PDR with a total column of $A_V = 6$~mags., are listed for the standard model parameters in Table~\ref{tab:pdrmodeloutput}. The column of C and C$^+$ are both approximately one-third that of the total column of gas-phase CO found though the PDR.

\begin{figure}
\includegraphics[angle=0,scale=0.7]{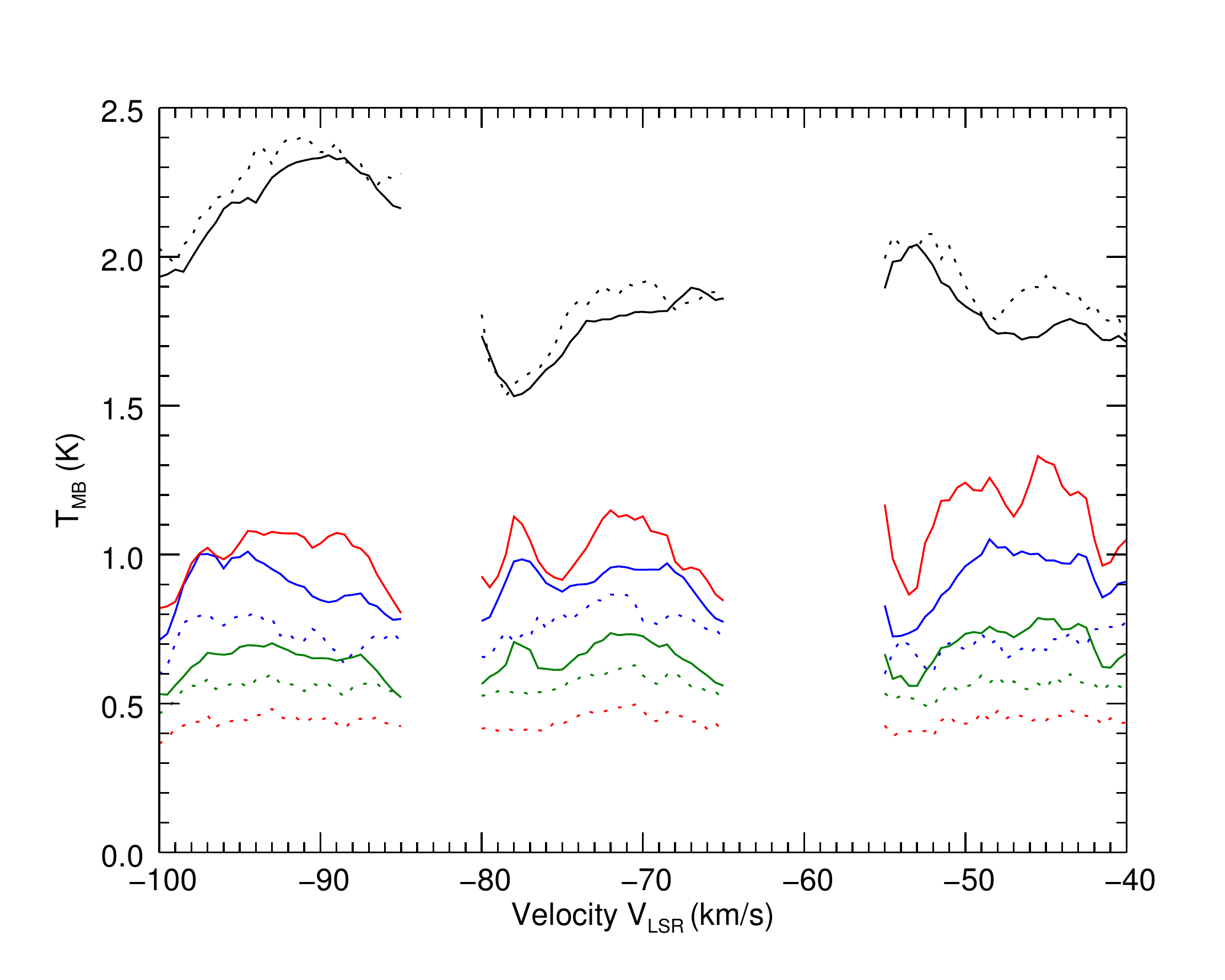}
\caption{Mean line brightness per velocity channel for the \coa/5 (blue), \cob\ (red), \ci\ (green) and \hi/150 (black) lines.  Solid curves are for voxels where \ci/\cob\ $< 1.0$ (i.e.\ ``normal'') and dotted for voxels where \ci/\cob\ $> 1.0$ (i.e.\ ``high'').  The velocity ranges shown are chosen for the three spiral arm crossings discussed in \S\ref{sec:obs} (i.e.\ $-100$ to $-85$~km/s, $-80$ to $-65$~km/s and $-55$ to $-40$~km/s, with the narrow filament being contained within the first of these). 
  The flux in a voxel needs to exceed the thresholds in Table~\ref{tab:voxels} to be included in this analysis.  Profiles are then averaged over all the voxels which exceed the thresholds, and placed into the appropriate line ratio range (i.e.\ ``high'', ``normal'') for display. Voxels that do not meet the threshold criteria are excluded from the averages shown here.  The normalization for each velocity channel is thus different than in Figs.~\ref{fig:profiles} and \ref{fig:avprofiles}. \label{fig:lineprofiles}}
\end{figure}

Comparison with the flux measurements in Fig.~\ref{fig:lineprofiles} shows that these model predictions are consistent with the data for \coa, \cob\ and \ci\ 2--1, if the average beam filling factor in the $2'$ voxel is $\sim 50\%$ (and taking into account the difference between the model 4~km/s FWHM and the voxel channel width of 0.7~km/s).  The corresponding \ci\ 1--0 and \cii\ fluxes are thus predicted to be of order 7 and 1\,K\,km/s, respectively, if observed with the same spectral and spatial resolution.

The models show that the \ci\ 2--1/\cob\ and \coa/\cob\ line ratios are constant for clouds with total extinctions $A_{V,{\rm tot}} > 2-4$ mags.\ (dependent on the particular model), with ratios of $\sim 1$ and $\sim 4$, respectively.  For lower $A_{V,{\rm tot}}$ these line ratios rise, as the relative line fluxes are sensitive to the depth of the PDR\@. Small $A_{V,{\rm tot}}$ corresponds to PDRs which are either relatively
thin, or for sight lines which are predominantly passing through the surface layers of PDRs. This latter case would also correspond to sight lines passing through the edges of molecular clouds rather than through their cores.  The ratios can rise to high values ($>3$ and $>15$, respectively) for small $A_{V,{\rm tot}}$, but then the corresponding line fluxes are themselves small as only a limited column of gas is being sampled.  No data with such high ratios was seen, though possibly with higher angular resolution (e.g.\ as with ALMA), when beam dilution of any such smaller clouds that might exist would be much reduced, then higher ratios might be found in some gas.  

\begin{figure}
\includegraphics[angle=0,scale=0.7]{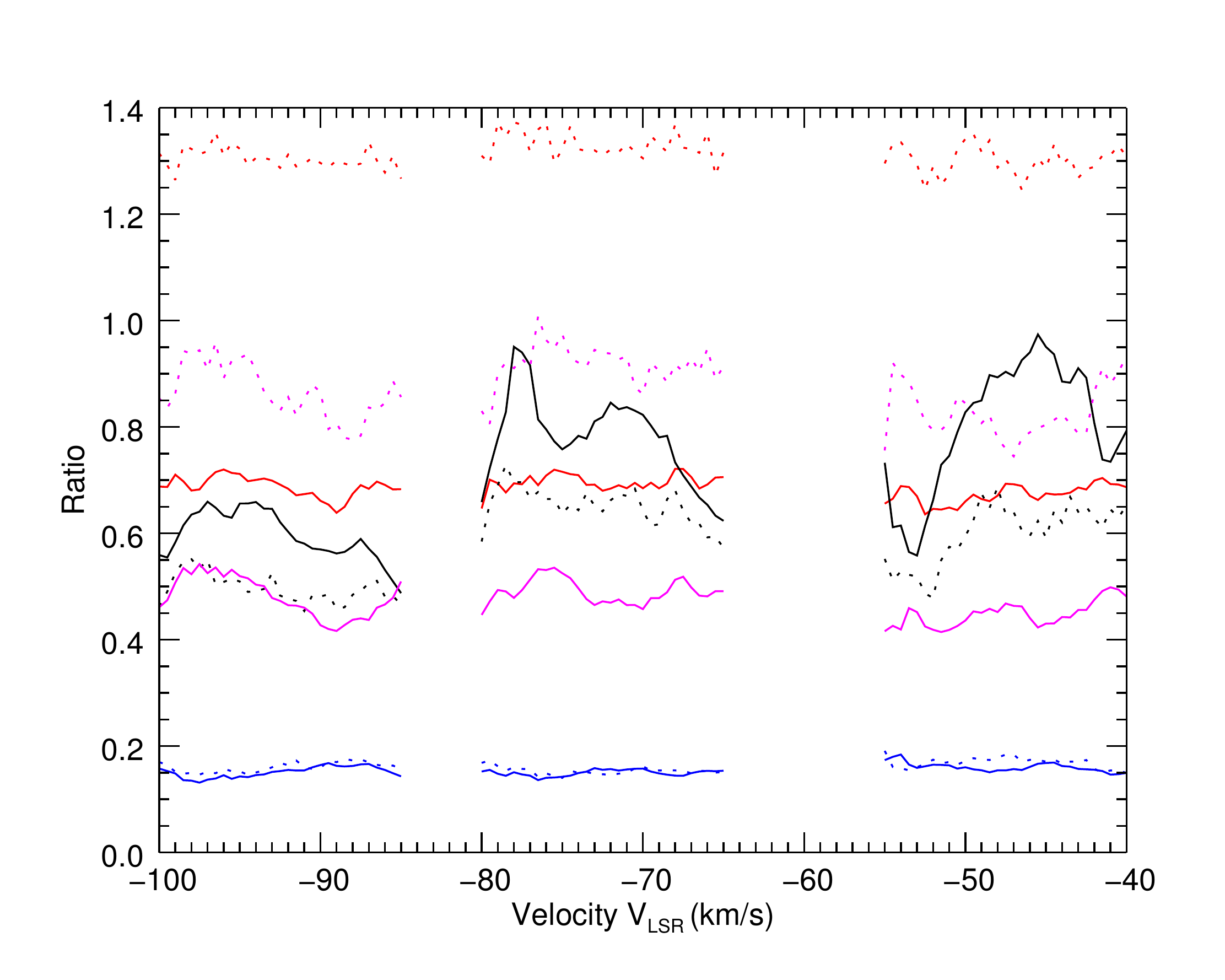}
\caption{As for Fig.~\ref{fig:lineprofiles}, but showing the averaged line ratios for (i)\ci/\cob\ (in red), (ii)\ci/\coa\ (in blue), (iii) [\coa/\cob] /10 (in magenta) and (iv) \ci/\hi\ $\times 100$ (in black), instead of the line intensities, for the same velocity ranges. As before, solid curves are for voxels where \ci/\cob\ $< 1.0$ (i.e.\ ``normal'') and dotted for voxels where \ci/\cob\ $> 1.0$ (i.e.\ ``high''). It is clear that for the ``high'' voxels the \ci/\cob\ and \coa/\cob\ ratios nearly double, \ci/\coa\ is unchanged and \ci/\hi\ decreases by $\sim 10\%$ (and noticeably more so in the velocity range of the narrow filament, $-80$ to $-76$~km/s). \label{fig:lineratios}}
\end{figure}

\begin{figure}
\hspace*{2cm}
\includegraphics[angle=0,scale=0.3]{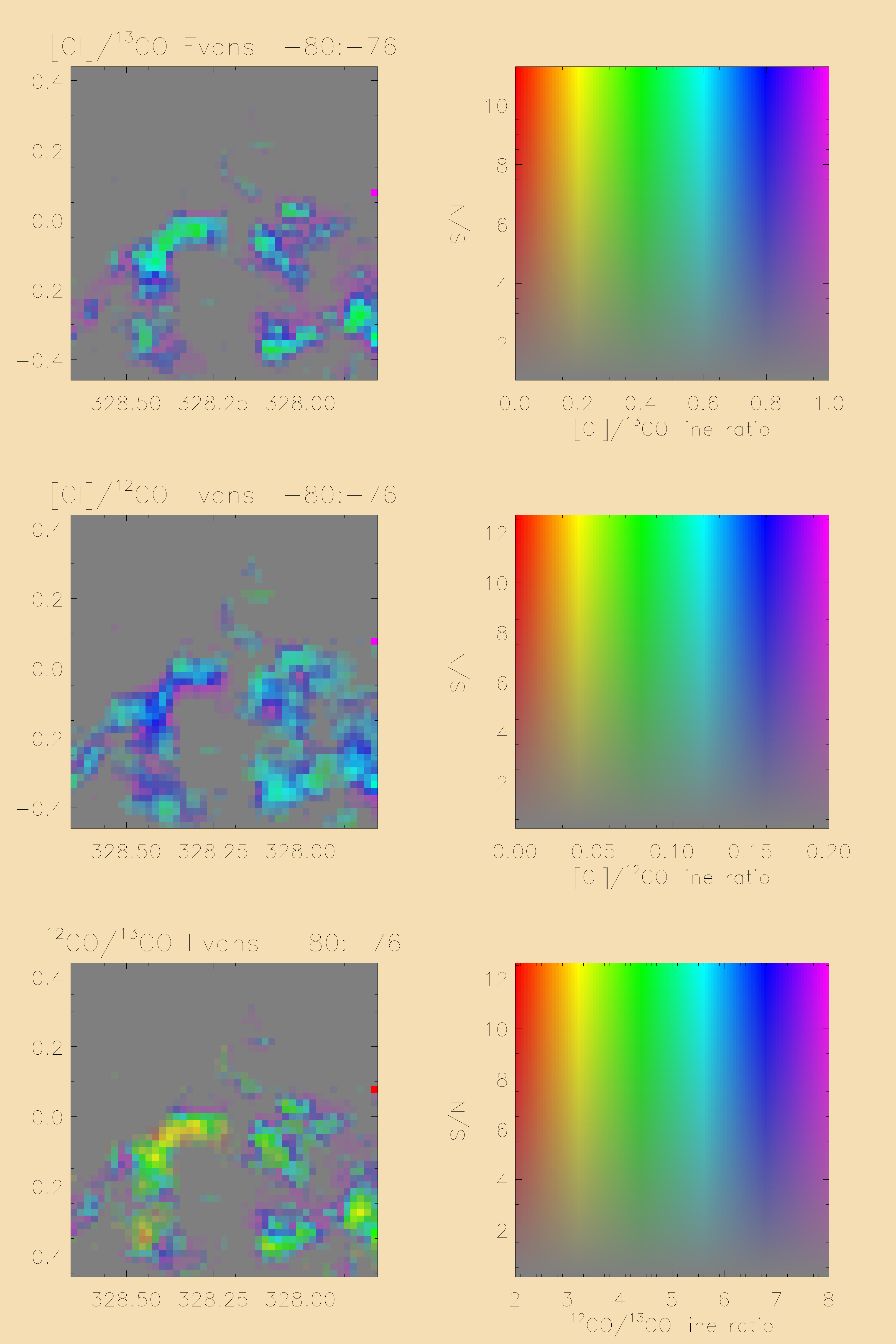}
\caption{Evans plots for the velocity range $-80$ to $-76$ km/s corresponding to the filament (left) and their corresponding 2D color tables (right). Shown are, from top to bottom, \ci/\cob, \ci/\coa\ and \coa/\cob.  Color (hue) denotes the value of the line ratio $x-$axis of the color table), and the saturation (intensity) its S/N, with low S/N pixels fading to grey ($y-$axis of the color table). \label{fig:evans_v8076}}
\end{figure}

\begin{figure}
\hspace*{2cm}
\includegraphics[angle=0,scale=0.3]{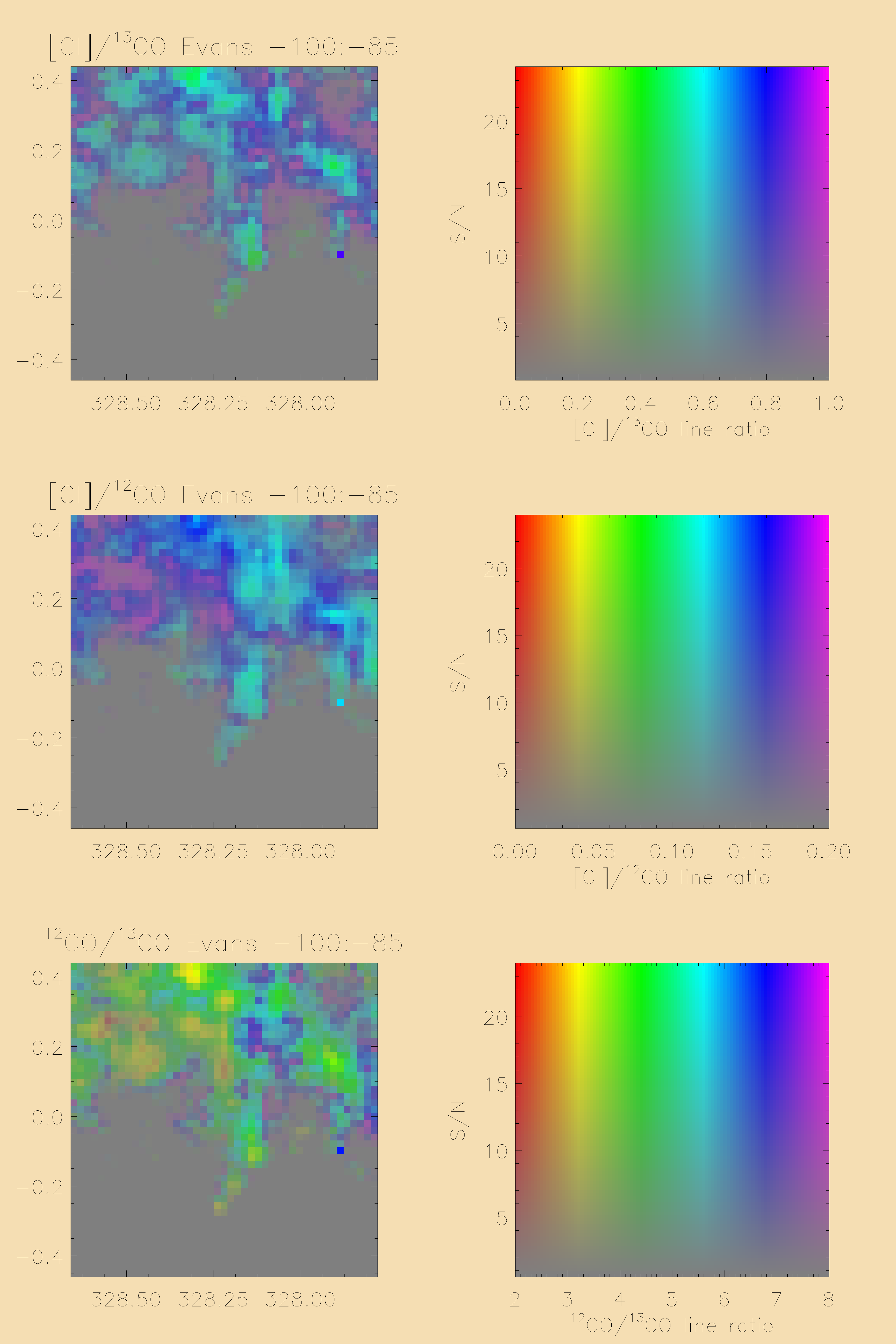}
\caption{Evans plots for the velocity range $-100$ to $-85$ km/s corresponding to the Norma near-spiral arm crossing (left) and their corresponding 2D color tables (right). Other details are as for Fig.~\ref{fig:evans_v8076}. \label{fig:evans_v100}}
\end{figure}

\begin{figure}
\includegraphics[angle=0,scale=0.8]{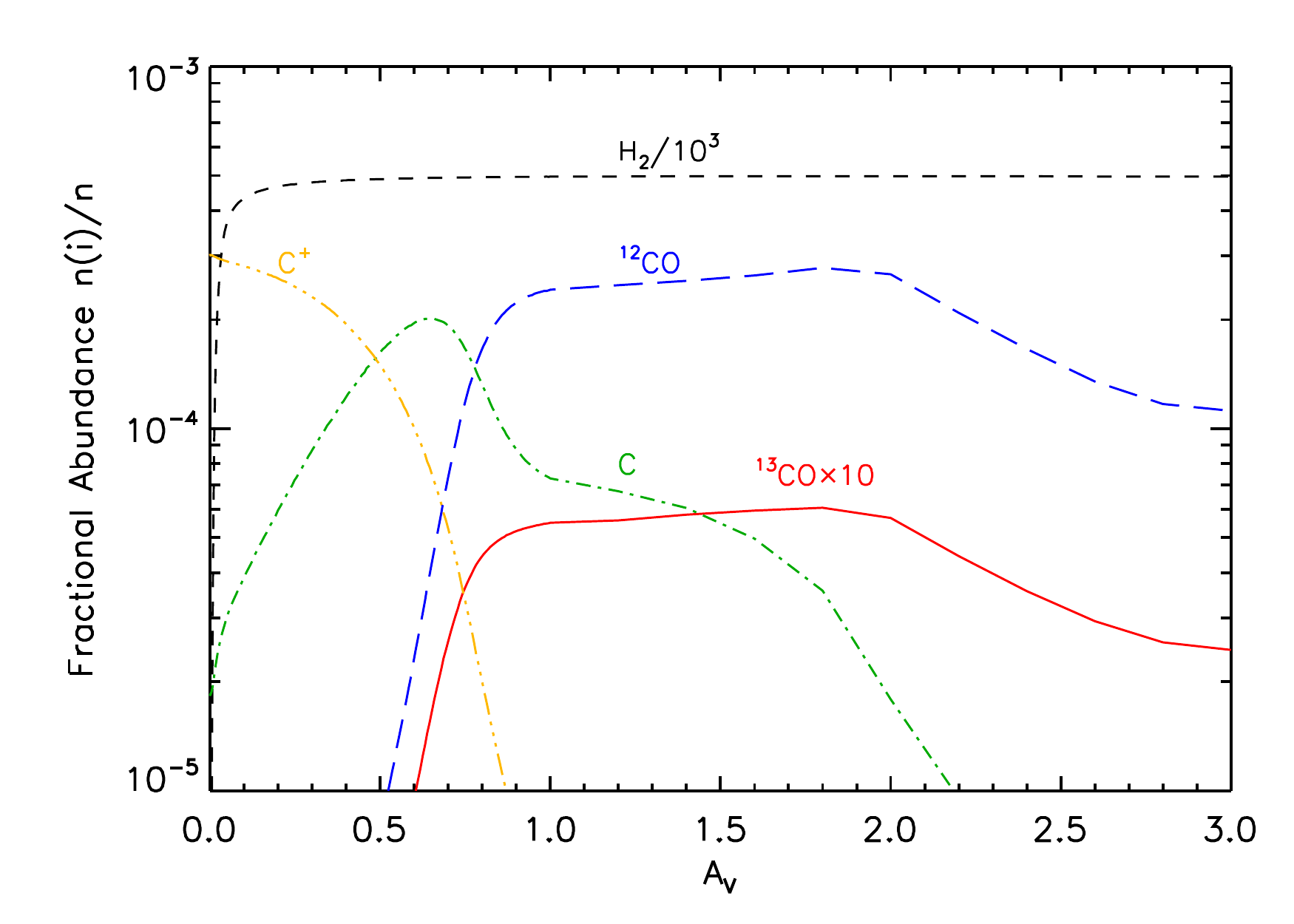}
\caption{Abundance of H$_2 / 10^3$ (black dash), C$^+$ (yellow dash-dot-dot), 
C (green dot-dash), \coa\   (blue long-dash), and \cob $\times 10$ (red solid),  as a function of optical depth $A_V$ (in magnitudes) for the standard cloud model. \label{fig:xh2covavplt}}
\end{figure}

\begin{figure}
\includegraphics[angle=0,scale=0.8]{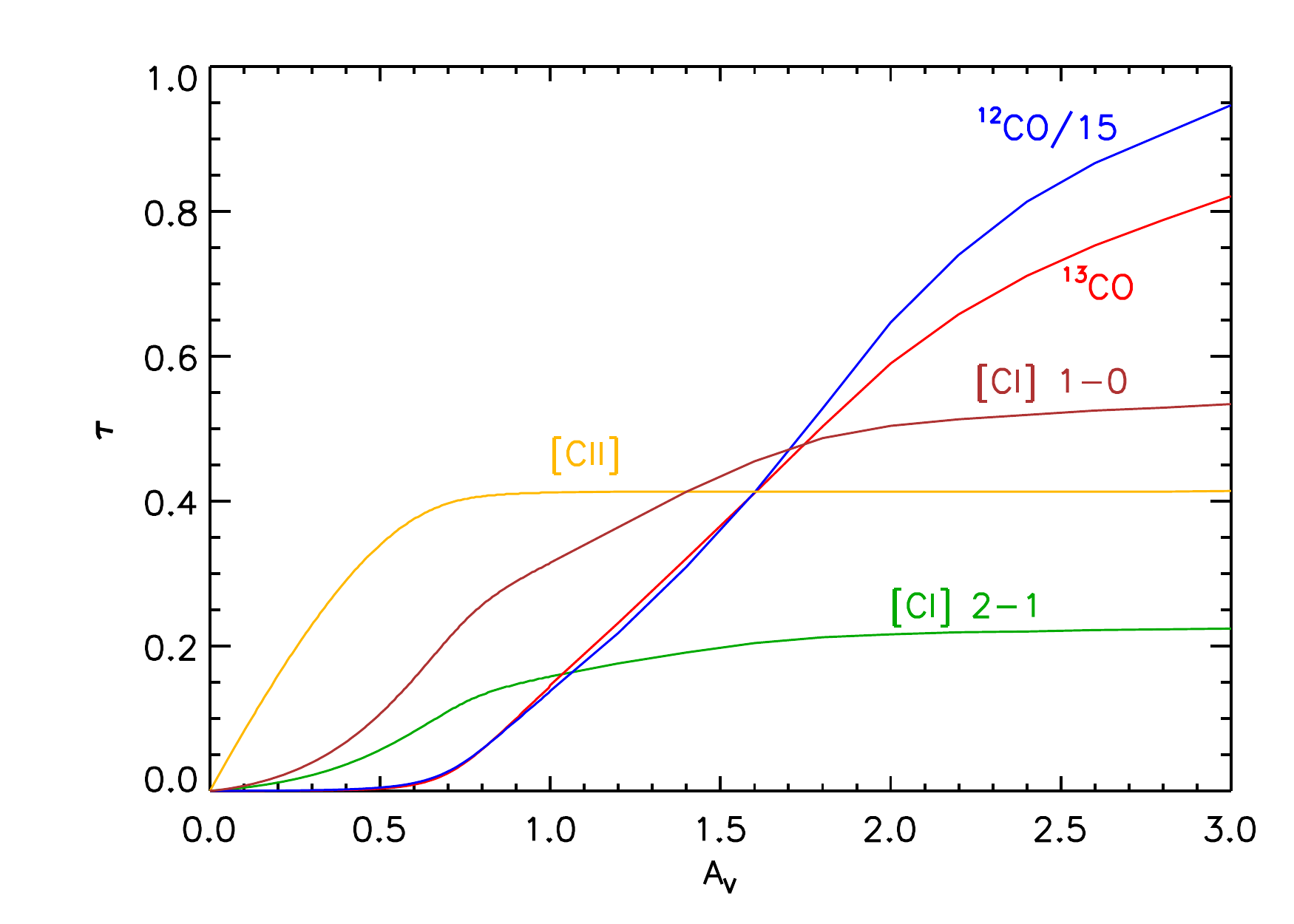}
\caption{Optical depth of the emission lines for  \ci\ 1--0 (brown) and  2--1  (green), \coa/15 (blue) and  \cob\ (red) 1--0, and \cii\ (yellow), as a function of $A_V$ for the standard model.   The total extinction through the cloud is $A_{V,{\rm tot}} = 6$ mags., and lines are plotted to the cloud center. The total optical depth in the line through the cloud is thus twice that shown at $A_V = 3$. \label{fig:tauci13covavplt}}
\end{figure}

\begin{figure}
\includegraphics[angle=0,scale=0.8]{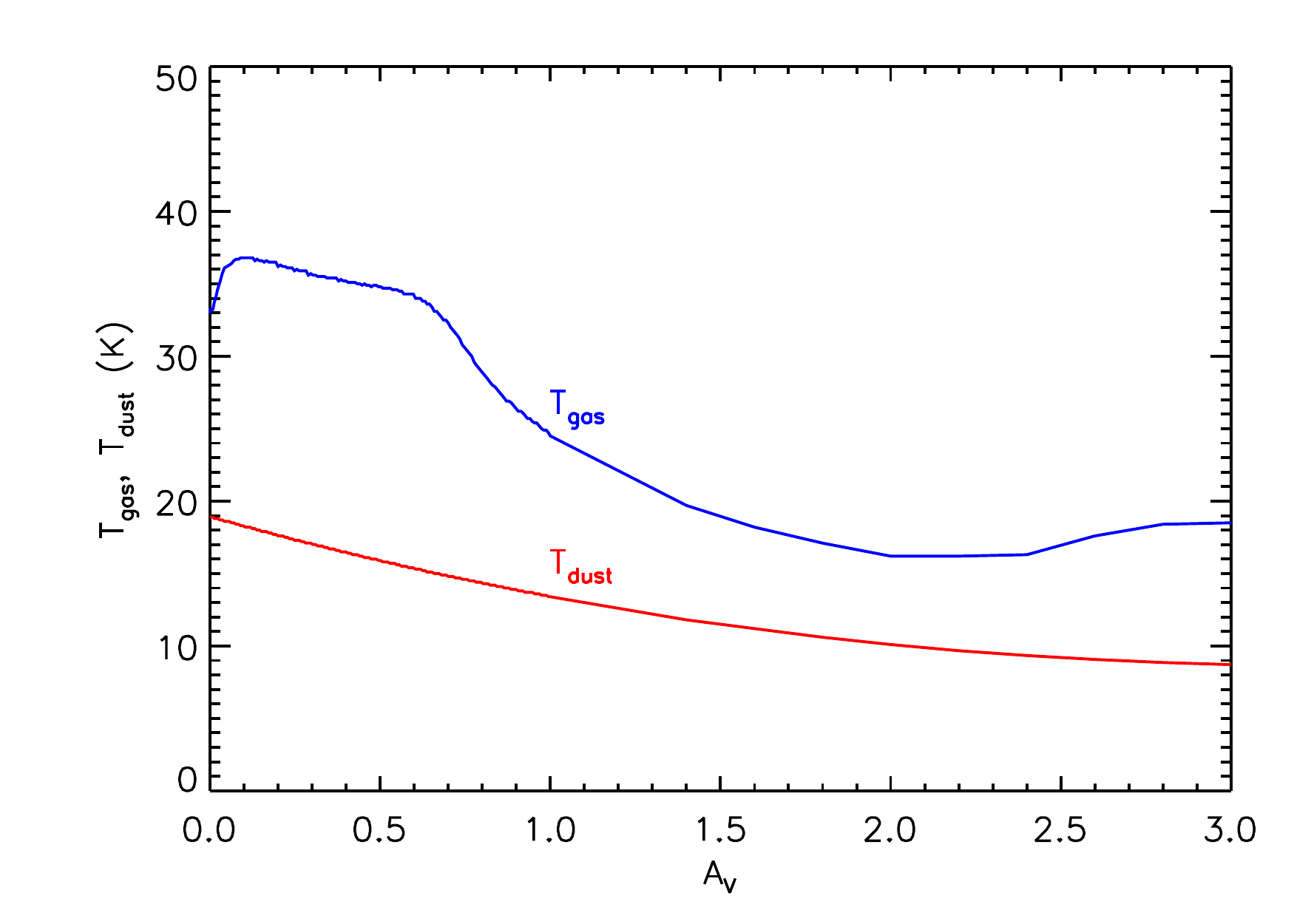}
\caption{The gas temperature, $T_{\rm gas}$, and dust temperature $T_{\rm dust}$  as a function of $A_V$ for the standard 
model. \label{fig:tgasplt2}}
\end{figure}

\begin{figure}
\includegraphics[angle=0,scale=0.8]{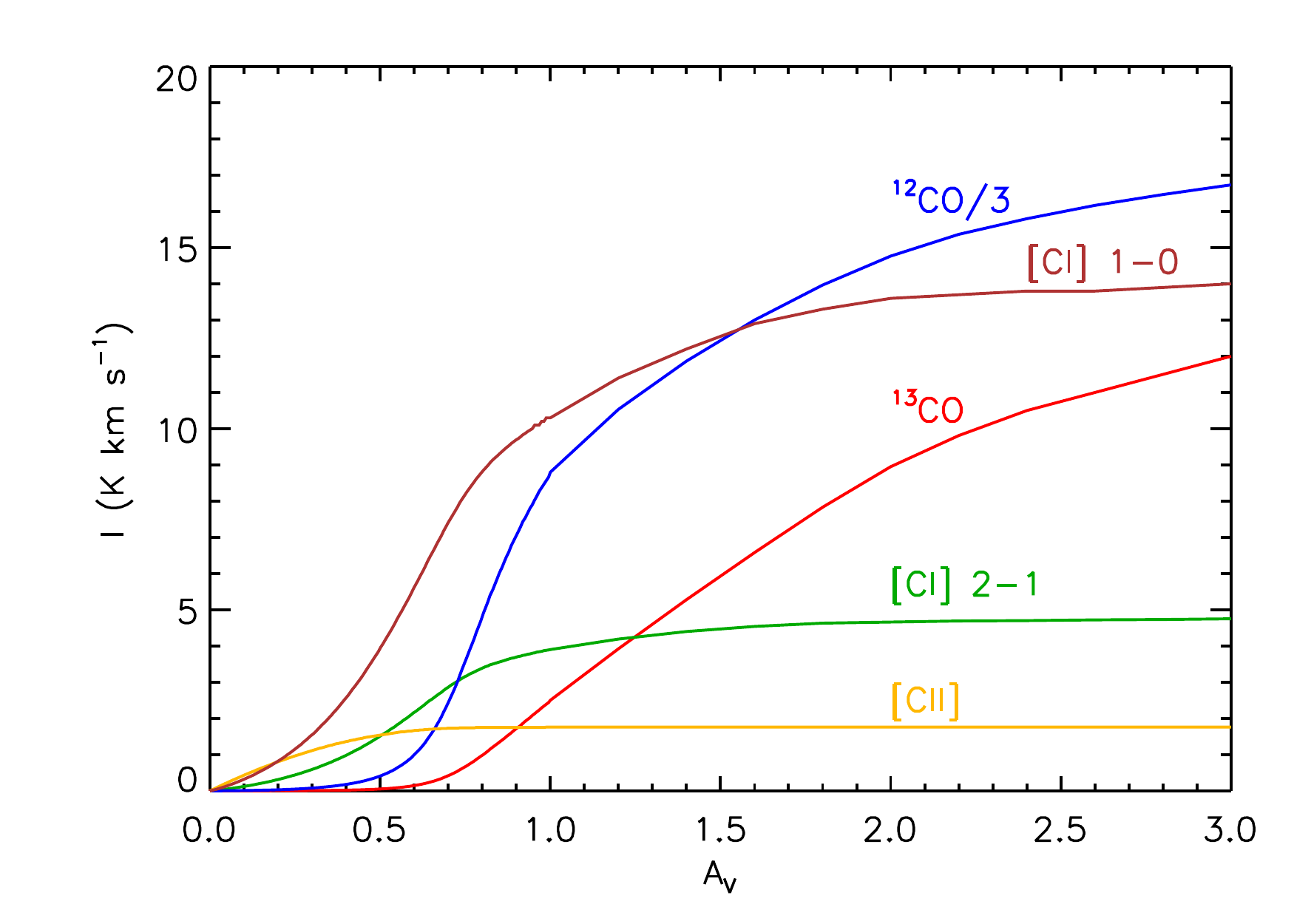}
\caption{Integrated intensities, in K km/s, of the emission lines for \coa/3 (blue), \cob\ (red) 1--0, \ci\ 1--0 (brown), \ci\ 2--1 (green) and \cii\ (yellow), as a function of optical extinction, $A_V$, into the PDR, for the standard model.   The total extinction in magnitudes through the cloud is $A_{V,{\rm tot}} = 6$. The intensity plotted is the total intensity emitted through one side of the cloud including the emission from both the near and far sides. \label{fig:ci13coTkmvAv}}
\end{figure}

\begin{figure}
\includegraphics[angle=0,scale=0.8]{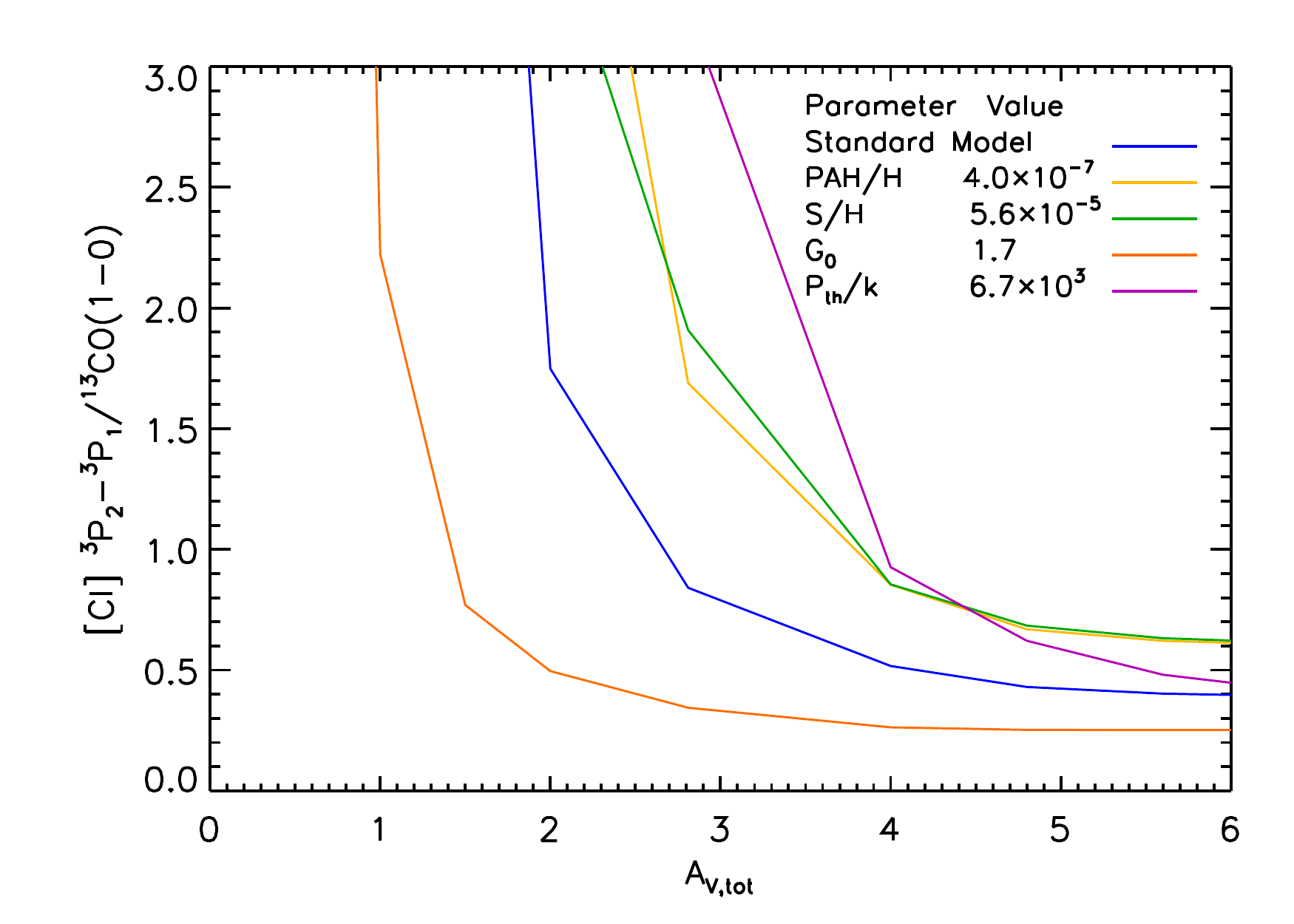}
\caption{The \ci\ 2--1 / \cob\ 1--0 line ratio as a function of total optical extinction through the cloud, $A_{V\rm ,tot}$, for a range of model parameter values, as indicated in the legend. 
\label{fig:ci13covAv3}}
\end{figure}

\begin{figure}
\includegraphics[angle=0,scale=0.8]{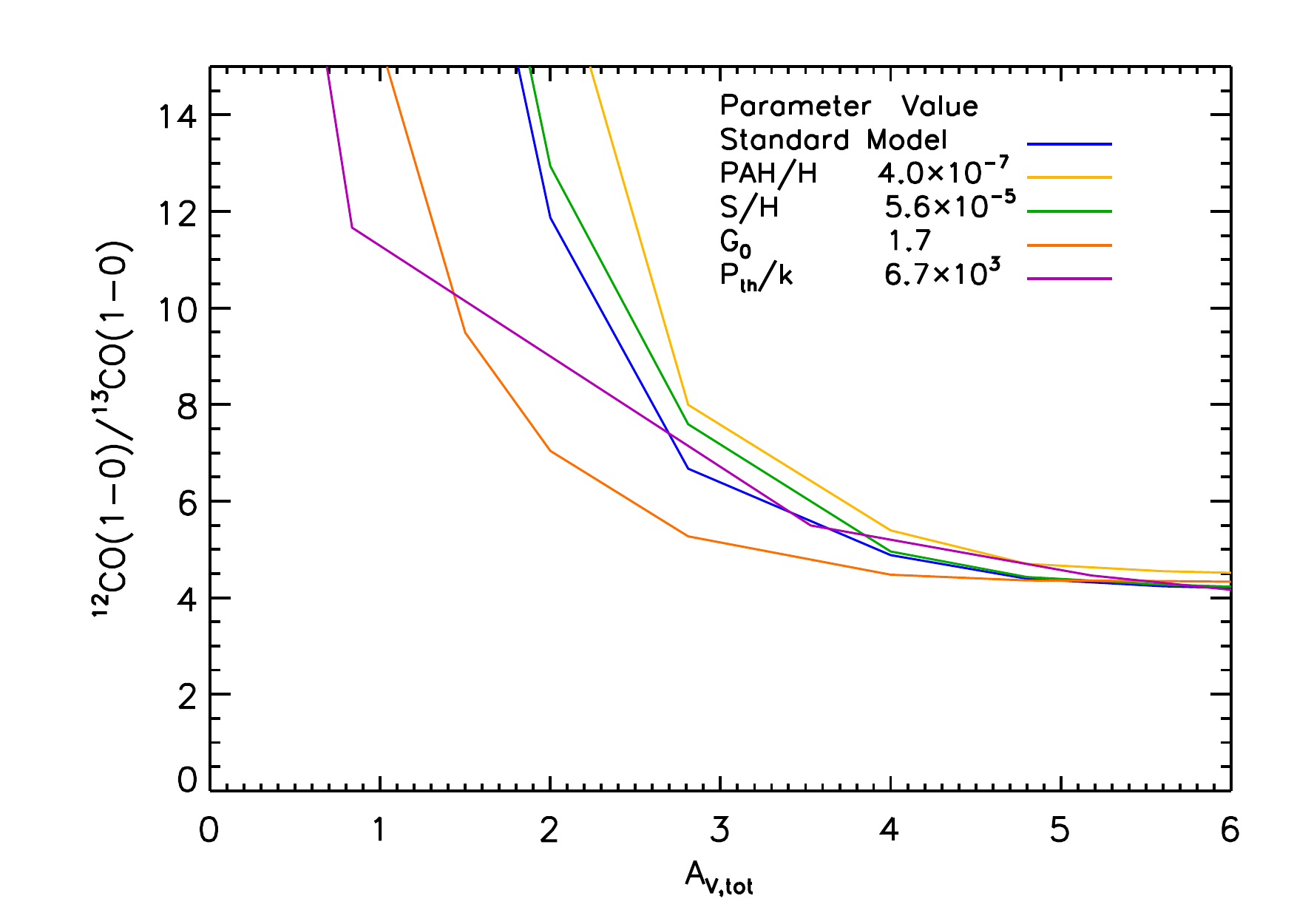}
\caption{The \coa/\cob\ J=1--0 line ratio as a function of total optical extinction through the cloud, $A_{V,{\rm tot}}$, for a variety of model parameters as indicated in the legend. \label{fig:co12co13vAv}}
\end{figure}

On the other hand, limb brightening would result in larger line fluxes.  Since the measured line fluxes are actually smaller when \ci/\cob\ is larger (see Figs.~\ref{fig:ratiovsflux} and \ref{fig:lineratios}), this implies that any limb brightening that occurs is overwhelmed by smaller beam filling factors. 

The typical \ci/\cob\ line ratios measured here are $\sim 0.7$, rising to $\sim 1.3$ (Fig.~\ref{fig:lineratios}).  The latter are found where the fluxes are, in general, weaker (Fig.~\ref{fig:ratiovsflux}).  While for individual pixels the effects of weak intrinsic flux or beam dilution cannot be disentangled, the modelling shows that if the \cob\ is intrinsically weak then the \ci/\cob\ ratio is higher (Fig.~\ref{fig:cico13vco13}).  This is consistent with the associated data being dominated by voxels with intrinsically weaker flux rather than by beam dilution.

\begin{figure}
\includegraphics[angle=0,scale=0.8]{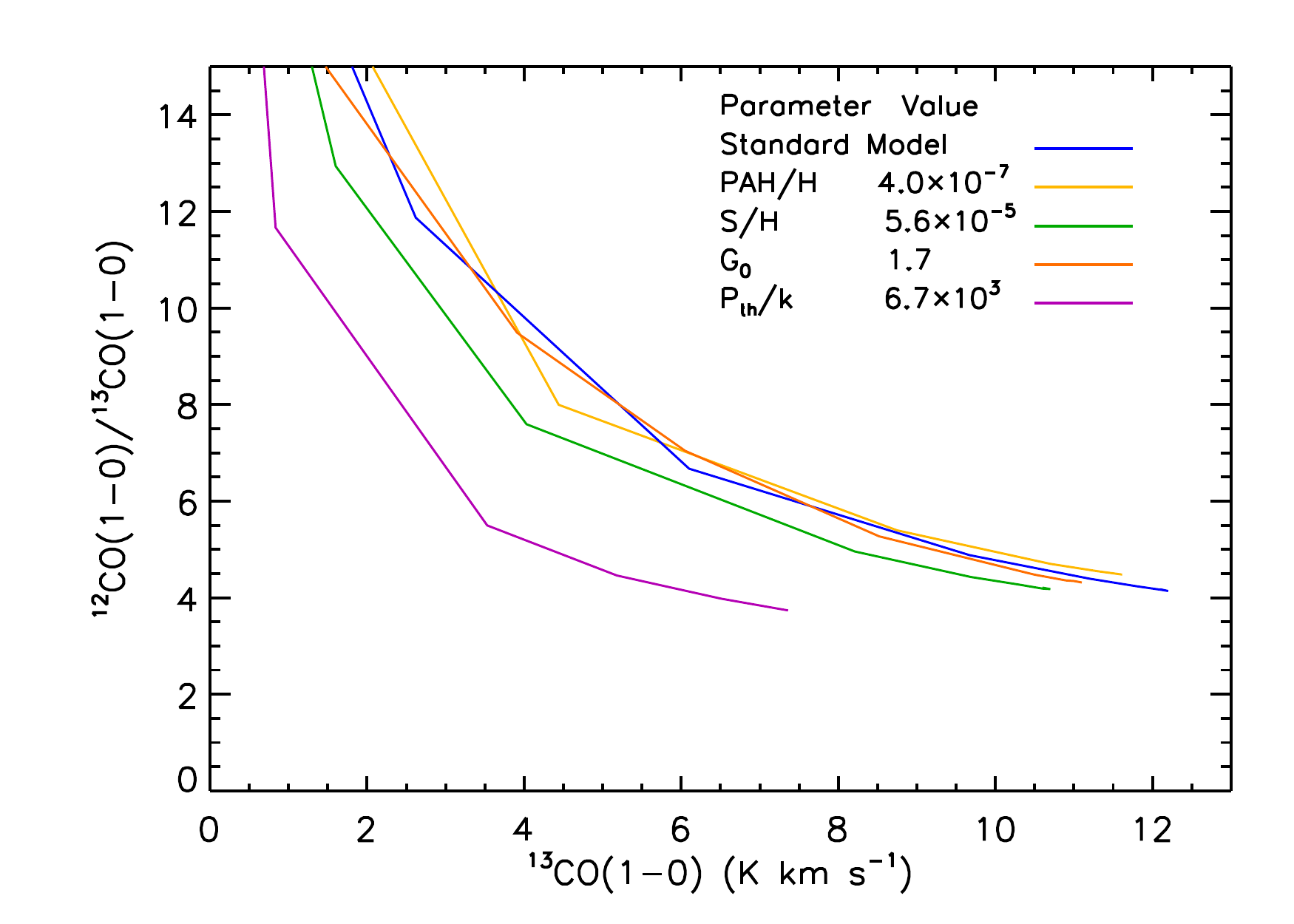}
\caption{\coa/\cob\ J=1--0 line ratio as a function  of the \cob\ line flux for a variety of model parameters, as indicated in the legend. \label{fig:co12co13vco13}}
\end{figure}

\begin{figure}
\includegraphics[angle=0,scale=0.8]{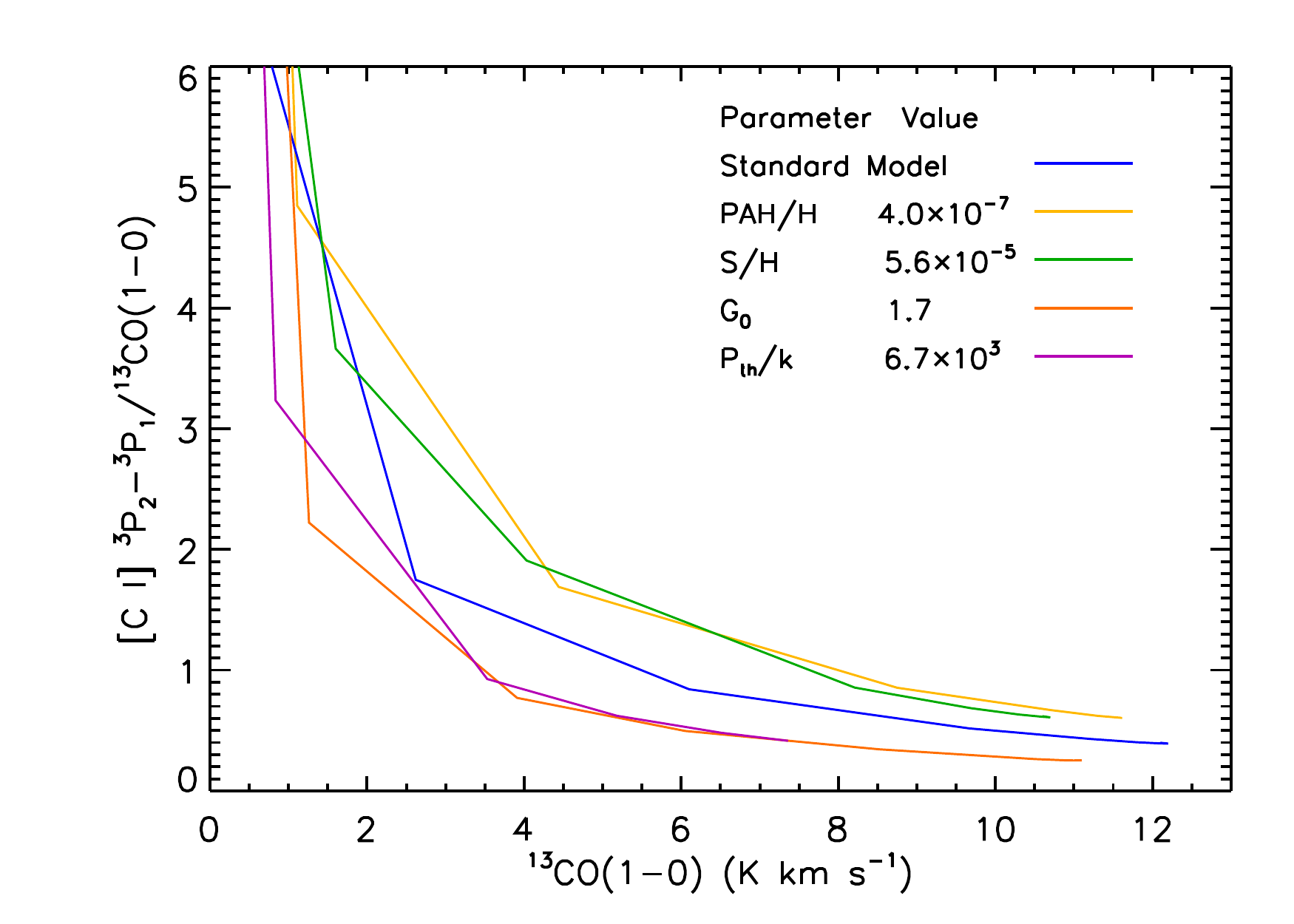}
\caption{\ci\ 2--1 / \cob\ line ratio as a function  of the \cob\ line flux for a variety of model parameters, as indicated in the legend. \label{fig:cico13vco13}}
\end{figure}

Typical \coa/\cob\ line ratios observed are $\sim 5$, rising up to $\sim 15$ (Fig.~\ref{fig:lineratios}). As for \ci/\cob, the latter generally occur where \cob\ fluxes are lower.  And similarly (Fig.~\ref{fig:co12co13vco13}), these values are consistent with the higher ratios being associated with sight lines passing through lower column density gas (and/or just the front surfaces of PDRs).

To summarize, the line fluxes for \coa, \cob\ and \ci\ presented here for the G328 sightline of the Galaxy are consistent with being produced in low density ($n_{\rm H_2} \sim 600$\,cm$^{-3}$) photodissociation regions being excited by far--UV radiation fields with strengths around the typical average interstellar value ($G_0 \sim 1.7 - 5.1$).  Larger line ratios of \ci/\cob\ are found, in general, associated with smaller values of the \cob\ line flux.  These smaller line fluxes are, in turn, generally associated with being emitted by gas with lower intrinsic line fluxes, rather than arising in sight lines with more beam dilution (which is found to average around 50\%).  Such regions are associated with the front surfaces of PDRs, within an optical extinction of 2 mags.\ from the atomic hydrogen interface.  Here carbon can be found in ionized (C$^+$), neutral (C) or molecular (CO) form, rather than being dominated by CO at greater depths.  Such regions are also associated with dark molecular gas, molecular clouds (i.e.\ H$_2$) where CO does not provide the dominant tracer for the gas.  

\section{Summary}
\label{sec:summary}
We have presented data cubes showing the distribution of the \ci\ 2--1, \coa, \cob\ and \coc\ 1--0, and HI line emission, over a $\sim 1^{\circ}$ region of the Galactic plane along a sightline towards G328, with angular and spectral resolutions of $\sim 1'$ and $\sim 1$~km/s.  The \ci\ data comes from a new telescope, HEAT, sited on the summit of the Antarctic plateau at the driest location on the Earth, where the THz atmospheric windows are opened for observation.  The CO data was taken with the Mopra telescope, and the HI data with the Parkes and ATCA telescopes, all located in Australia.

Complex morphology is evident in all species, with the \ci\ and CO line emission extending over 120 km/s in extent, and arising principally from molecular cloud complexes in three spiral arm crossings along the sight line.  The distribution in these atomic and molecular species is very similar in both angular and spectral dimensions, and is encompassed by more extensive HI line emission.  However close examination of the \ci\ and CO emission shows the latter to be slightly more extended spatially.  

Average line ratios for \coa/\ci, \cob/\ci\ and \coa/\cob\ are found to $6.8 \pm 1.6, 1.4 \pm 0.5 \, {\rm and} \, 5.5 \pm 2.5$, respectively, across all voxels in the data cubes when resampled to the same resolution.  These ratios are relatively constant with velocity, across the three spiral arms crossed along the sight line.  However, at the edges of the emission features the \ci/\cob\ and \coa/\cob\ ratios are typically found to be 50\% larger, with \ci/\hi\ ratios around 10\% lower.  This is attributed to relatively more C than CO at the edges of molecular clouds, with the \cob\ intensity providing the best measure for the column of CO, rather than the optically thick \coa.  On the other hand, the relative amount of atomic gas is seen to rise at the edges of the clouds.

PDR models were constructed for diffuse molecular gas exposed to average interstellar radiation fields to explore the behaviour of the \coa, \cob, \ci\ 1--0 and 2--1, and \cii\ line emission with increasing extinction, $A_V$, into molecular clouds from their front (atomic) surface. Charge exchange reactions between C$^+$ and both S and PAHs were also found to be important to match the data, and require significant depletions of the S and PAHs. The models then reproduce the broad behaviour seen in the data, in particular the line fluxes (with average beam filling factors of $\sim 50$\%) and the increased line ratios for \ci/\cob\ and \coa/\cob\ in the surface layers, i.e.\ when $A_V < 2$\,mags.\@ The highest values of the \ci/\cob\ ratio measured require sightlines passing through clouds that have extinctions $A_V < 1$ from their exciting sources.  These may arise from sightlines that only pass through the edges of clouds, and/or small clouds whose total column density extends for $A_V \sim 1$. The \cob\ and \ci\ lines remain marginally optically thin through the PDR, and so provide direct probes of the amount of CO and C present, respectively.  

Roughly one-third of the molecular gas along the sightline is estimated by these models to be associated with dark molecular gas, and about two-thirds in total when accounting for a spherical geometry.  This is gas where CO line emission does not provide a full tracer for the column of molecular gas that is present.  Considering the column of cold atomic gas in the front surface of the PDR, we also estimate that it contains $\sim 10\%$ of the total atomic gas along each sight line (the rest being in atomic gas clouds).  Of the atomic gas within the PDR, $\sim 15\%$ of it is dark, i.e.\ the fraction not measured through use of the standard optically thin assumption for determining column densities.  This is only $\sim 1\%$ of the total atomic column along sight line, however.

The modelling would also be aided by measurements of the \ci\ 1--0 line emission at 492 GHz, whose ratio with the \ci\ 2--1 line we have observed at 809\,GHz provides a probe that is sensitive to the gas temperature, predicted here to be $\sim 30$\,K for $A_V < 1$.  Such measurements could be made from the Atacama plateau in Chile, where the the superior angular resolution obtainable with larger telescopes such as Nanten2, APEX and ALMA would allow for a better comparison of the relative distributions of the C and CO, and so in determining more clearly the form and extent of the carbon envelope around molecular clouds.

A full quantification of the amount of dark molecular present also requires measurement of the \cii\ line emission, which is confined within $A_V < 0.5$ of the PDR surface. Some airborne and space measurements currently for such \cii\ now exist \citep[e.g.][]{2014A&A...570A.121P, Velu2015}.  We provide predictions of its intensity and distribution with extinction into molecular clouds, to aid in their interpretation and for future observations made with THz facilities in Antarctica, such as the STO--2 long duration balloon-borne telescope scheduled for launch in 2016 and China's proposed 5m DATE5 telescope for Dome A on the summit of the Antarctic plateau.

\acknowledgments
\section*{Acknowledgements}
Funding for the HEAT telescope is provided by the US National Science Foundation under grant number PLR-0944335.  PLATO--R was funded by Astronomy Australia Limited, as well as the University of New South Wales, as an initiative of the Australian Government being conducted as part of the Super Science Initiative and financed from the Education Investment Fund.  Support for the UNSW program in Antarctica is also provided by the Australian Antarctic Division.  Logistical support for HEAT and PLATO--R is provided by the United States Antarctic Program.  

The Mopra radio telescope is part of the Australia Telescope National Facility.  Operations support was provided by the University of New South Wales and the University of Adelaide.   The UNSW Digital Filter Bank used for the observations with Mopra  was provided with financial support from the Australian Research Council (ARC), UNSW, Sydney and Monash universities.  We also acknowledge ARC support through Discovery Project DP120101585. M.G.W.\@ and D.J.H.\@ were supported in part by NSF grant AST--1411827.

M.G.B.\@ thanks the Dublin Institute for Advanced Studies (DIAS) in Ireland and the University of Leeds in the UK for their hospitality, during which much of the analysis for this paper was undertaken.  



{\it Facilities:} \facility{Mopra}, \facility{HEAT}, \facility{Parkes}, \facility{ATCA}.

\clearpage

\section*{Tables}
\begin{deluxetable}{ccccccc}
\tablecaption{Statistics on the Voxels used for the Histogram Analysis \label{tab:voxels}}
\tablehead{Line & Number & n$\sigma$ & Threshold (K) & Number & n$\sigma$ & Threshold (K)} 
\startdata
~\coa\  & 204,350   &  1   &  0.9 & 90,059 & 3 & 3\\
~\cob\  & 148,623   &  1    & 0.3 & 38,379 & 3 & 0.8 \\
~\coc\  & 91,875     &  1     & 0.2 &  2,783 & 3 & 0.6 \\
~\ci\   & 83,857     &  3     & 0.4  \\
HI       & 457,746    & Arbitrary    & 40 \\
\enddata
\tablecomments{Number of voxels (i.e.\ pixel-velocity channels)  that pass the threshold criteria used for the histogram ratio analysis (see \S\ref{sec:histograms}), together with the number of standard deviations ($\sigma$) these correspond to and the corresponding threshold values (in K).  For the three CO lines, the corresponding number of voxels that pass the threshold criteria if $3 \sigma$ is used instead of $1 \sigma$ is also listed.  The HI limit is arbitrary since there are no line-free channels to estimate the standard deviation from.  However, the S/N in the HI is very much higher than for the other lines, so this limit has little effect on the results.}
\end{deluxetable}

\clearpage
\begin{deluxetable}{ccccccccccc}
\tablecaption{Line Ratio Statistics \label{tab:voxelstats}}
\tablehead{Ratio & \#  & Mean & Med. & Max. & $\sigma$ & \#  & Mean & Med. & Max. & $\sigma$ \\ 
Pair & \multicolumn{5}{c}{$1 \sigma$ threshold for CO} & \multicolumn{5}{c}{$3 \sigma$ threshold for CO}}
\startdata
~\coa/\ci     & 83,545   &   6.8   &  6.7   &  18    & 1.6     & 72,112  & 7.1   & 7.0   & 18    & 1.5 \\
~ \ci/\coa    & 83,545   &  0.16  & 0.15  &  0.7   & 0.04   & 72,112  & 0.15 & 0.14 & 0.34 & 0.03 \\
~ \coa/\cob & 115,687 &  5.5    &  5.0   &  25    & 2.5     & 36,301  & 4.3   & 4.2   & 10    & 1.2 \\
~ \ci/\cob    & 74,063  &   0.84  &  0.76 & 3.0    & 0.33  & 36,751  & 0.63  & 0.61 & 1.6  & 0.14 \\
~ \cob/\ci    & 74,063  &  1.4     & 1.3    & 4.1   & 0.5     & 36,751 & 1.7     & 1.6   & 4.1  & 0.4 \\
~ \coa/\coc  & 46,750   &  10     &  8.7   &  62    & 6.1     & 1,715    & 7.8   & 7.0   & 22    & 3.2 \\
~ \coa/HI     & 203,197 &  0.03  & 0.02  & 0.34  & 0.02  & 89,828   & 0.05 & 0.04  & 0.34& 0.02 \\
~ \ci/HI        & 83,588  &  0.007 & 0.006& 0.047& 0.003 \\
~\cob/HI      & 142,961& 0.007  & 0.005& 0.12  & 0.006 & 38,181 & 0.007 & 0.012 & 0.12 & 0.008 \\
\enddata
\tablecomments{Statistics for the voxels which exceed the threshold limits in Table~\ref{tab:voxels} for each line pair.  Columns 2--6 refer to the $1 \sigma$ threshold for the CO lines, columns 7--11 to the $3 \sigma$ threshold.  \# is the number of voxels that meet the threshold limits. Both the mean and median values are shown, as well as the maximum value measured for the ratio and the standard distribution of the distribution. For \ci/HI only one set of threshold limits has been used, hence the values are only tabulated once. Fig.~\ref{fig:histograms} shows the histograms for the corresponding line ratio distributions. } 
\end{deluxetable}

\clearpage
\begin{deluxetable}{lc}
\tablewidth{0pt}
\tablecaption{PDR Model Inputs\label{tab:pdrmodel}}
\tablehead{
\multicolumn{2}{c}{Input Parameters for the Standard Model}\\
{Parameter} & \colhead{Value}}
\startdata
$G_0$ & 5.1 Habings \\
$A_{V,{\rm tot}}$ & 6 mags. \\
& (unless otherwise specified) \\ 
$P_{\rm th} / k$ & $\rm 2.0 \times 10^4$  K  ${\rm cm^{-3}}$ \\
$\Delta V$ (FWHM) & 4.0 km  s$^{-1}$  \\
$\zeta_{\rm crp}$  & $2.0 \times 10^{-16}$ ${\rm s^{-1}}$ \\
${\rm [^{12}C/^{13}C}]$ isotopologue ratio & $50$ \\
$\rm [C/H]$ abundance  & $3.2\times 10^{-4}$ \\
$\rm [S/H]$ abundance  & $7.0\times 10^{-6}$ \\
$\rm [PAH/H]$ abundance  & $2.0\times 10^{-7}$ \\
\enddata
\tablewidth{450pt}
\tablecomments{
Parameters for the 2-sided PDR models (i.e.\ the model is 1D with radiation incident from both sides).  
$G_0$ is the radiation field strength in free space in units of the Habing field ($\rm 1.6 \times 10^{-3} \, 
ergs \, s^{-1} \, cm^{-2}$; a representative value for the average interstellar radiation field). $P_{th}/k$ 
is the thermal pressure in K\,cm$^{-3}$. $\Delta V$  is the FWHM line width in km s$^{-1}$.  $\zeta_{\rm crp}$ 
is the primary cosmic ray ionization rate per H nucleus. }
\end{deluxetable}

\clearpage
\begin{deluxetable}{lc}
\tablewidth{0pt}
\tablecaption{PDR Model Outputs\label{tab:pdrmodeloutput}}
\tablehead{
\multicolumn{2}{c}{Output Predictions from the Standard Model}\\
{Parameter} & \colhead{Value}}
\startdata
$I(^{12}$CO 1--0) & 50 K km/s\\
$I(^{13}$CO 1--0) & 12 K km/s \\
$I$([CI] 1--0) & 14 K km/s \\
$I$([CI] 2--1) & 5 K km/s \\
$I$([CII]) & 2 K km/s \\
$I$(HI)  & 32 K km/s \\
N($^{12}$CO)  & $\rm 9.5 \times 10^{17}\,cm^{-2}$ \\
N($^{13}$CO)  & $\rm 2.1 \times 10^{16}\,cm^{-2}$ \\
N(C)  & $\rm 3.3 \times 10^{17}\,cm^{-2}$ \\
N(C$^+$)  & $\rm 2.9 \times 10^{17}\,cm^{-2}$  \\ 
N(H)  & $\rm 7.1 \times 10^{19}\,cm^{-2}$  \\ 
$T_{\rm gas}$ & 18 K \\
$T_{\rm dust}$ & 9 K \\
$T_{\rm gas}(\tau_{\rm CO} = 1)$ & 28 K \\
$T_{\rm dust}(\tau_{\rm CO} = 1)$ & 14 K \\
$n_{\rm H_2}(\tau_{\rm CO} = 1)$ & 580 cm$^{-3}$ \\
\enddata
\tablewidth{450pt}
\tablecomments{
Output parameters from the standard model of Table~\ref{tab:pdrmodel} are for a PDR extending to an optical extinction $A_{V,\rm{tot}} = 6$ mags.\ (i.e.\ $A_V = 3$ to its center).    Column densities (in cm$^{-2}$),  temperatures (in K) and line fluxes (in K km/s)  are as per Figs.~\ref{fig:xh2covavplt}, ~\ref{fig:tgasplt2} and \ref{fig:ci13coTkmvAv}, with the latter including the contributions from the near and far sides of the cloud. Also given are the temperatures and H$_2$ density when $\tau_{\rm CO}=1$. The fluxes are for the model FWHM of 4 km/s, and so need to be divided by 6 to be directly compared with the data, for which the voxel width is 0.7~km/s. }
\end{deluxetable}

\clearpage
\section*{Figures}

\clearpage

\clearpage

\clearpage

\clearpage


\clearpage

\clearpage


\clearpage

\clearpage

\clearpage

\clearpage


\clearpage

\clearpage

\clearpage

\clearpage

\clearpage

\clearpage

\clearpage

\clearpage

\clearpage
\appendix
\renewcommand\thefigure{\thesection.\arabic{figure}} 
\setcounter{figure}{0}    
\section{Appendix}
\begin{figure}
\includegraphics[angle=0,scale=0.33]{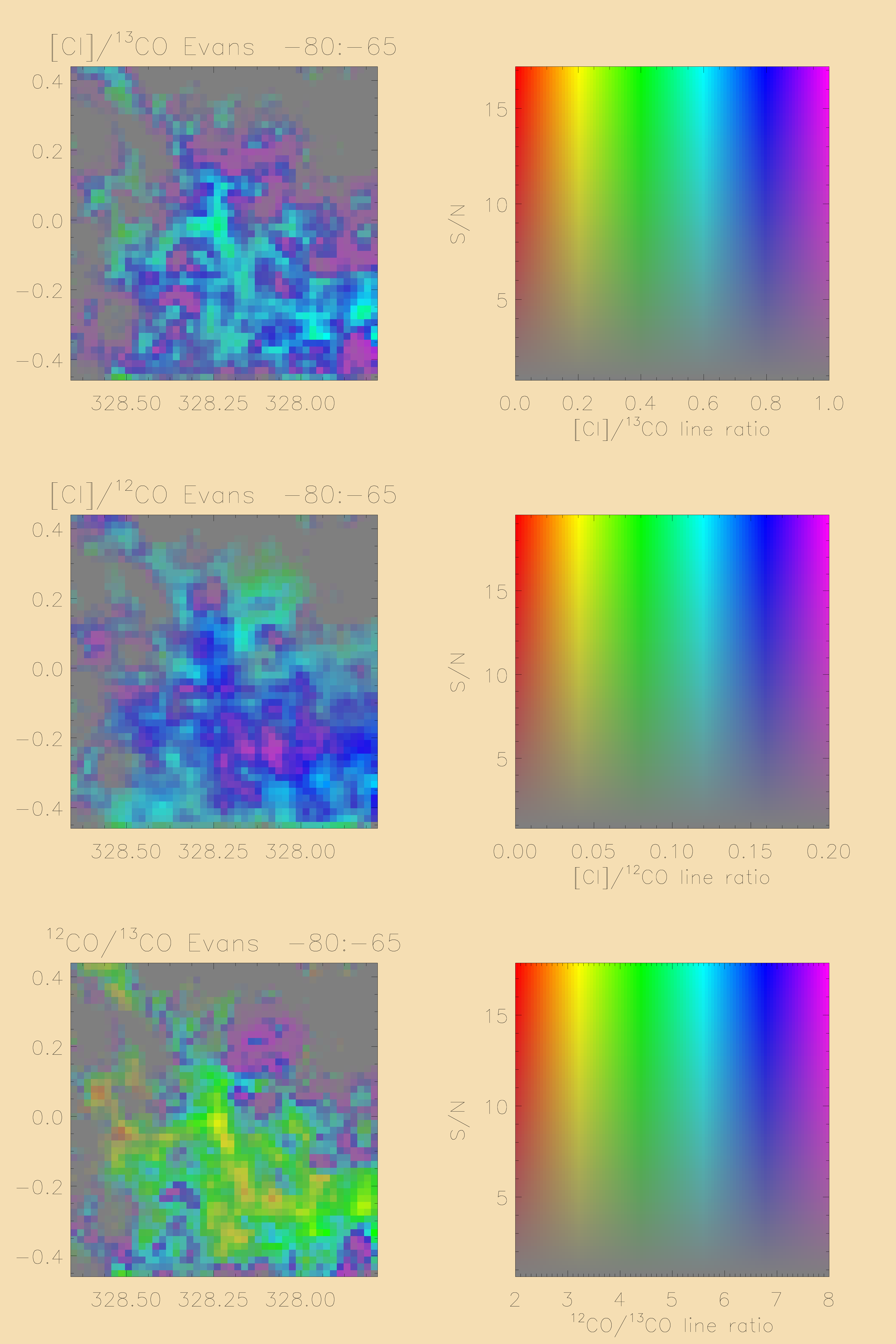}
\caption{Evans plots for the velocity range $-80$ to $-65$ km/s (left) and their corresponding 2D color tables (right). Shown are  [CI]/13CO, [CI]/12CO and 12CO/13CO.  Color (hue) denotes the value of the line ratio, and the saturation (intensity) its S/N.  This velocity range corresponds to the Norma far-spiral arm crossing. \label{fig:evans_v80}}
\end{figure}

\begin{figure}
\includegraphics[angle=0,scale=0.33]{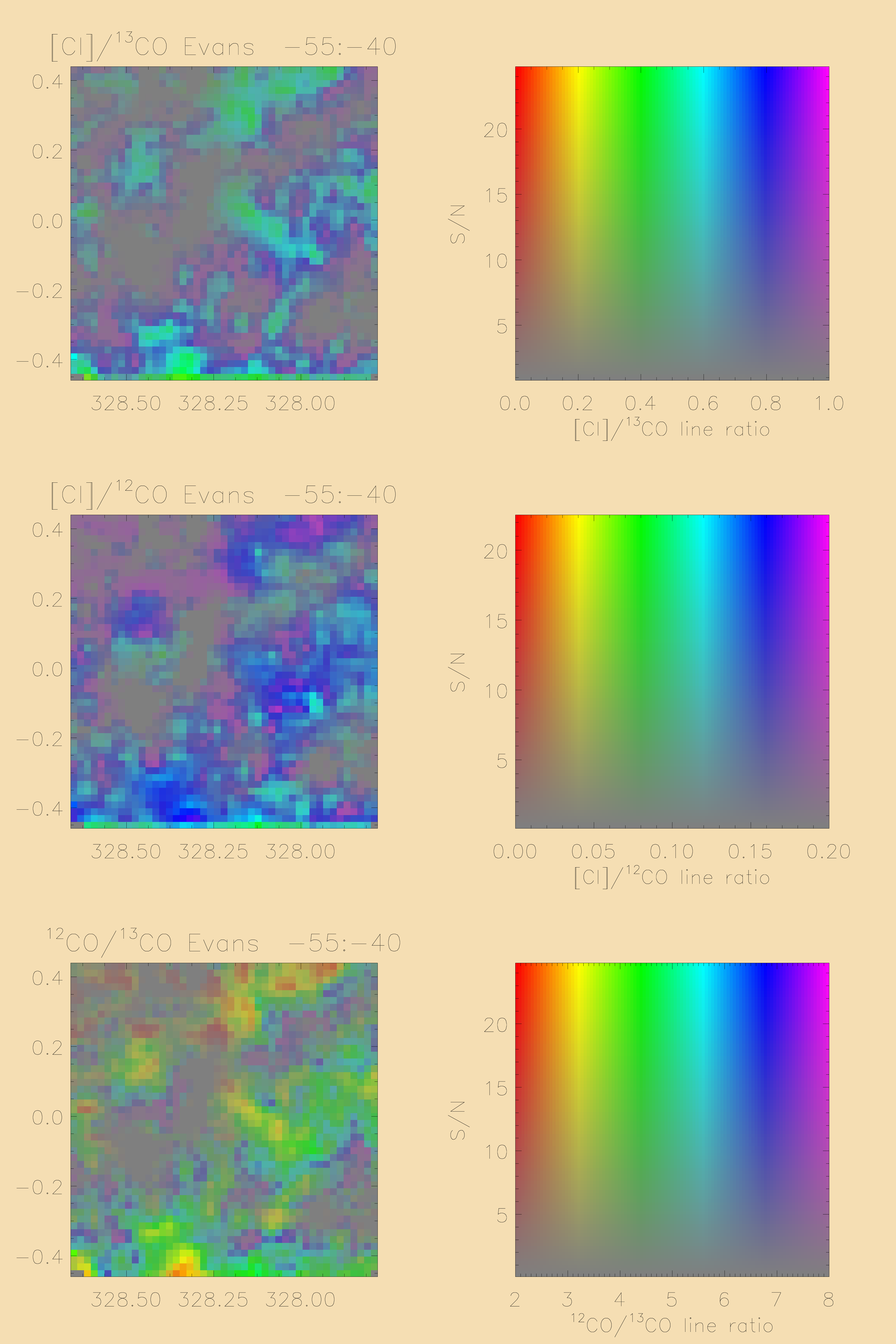}
\caption{Evans plots for the velocity range $-55$ to $-40$ km/s (left) and their corresponding 2D color tables (right). Shown are  [CI]/13CO, [CI]/12CO and 12CO/13CO.  Color (hue) denotes the value of the line ratio, and the saturation (intensity) its S/N\@. This velocity range corresponds to the Scutum-Crux near-spiral arm crossing. \label{fig:evans_v55}}
\end{figure}

Two Evans plots are shown in Figs.~\ref{fig:evans_v80} -- \ref{fig:evans_v55} for the velocity ranges of $\rm (-80,-65)$ and  $(-55,-40)$~km/s; associated with Norma far- and Scutum-Crux near- spiral arm crossings.   For each velocity range three plots are displayed, \ci/\cob, \ci/\coa\ and \coa/\cob, so the behaviour of all three lines can be compared.   The Evans plots are on the left and their corresponding 2D color tables on the right.

\clearpage

\end{document}